%% file: main.tex
\documentclass[10pt,twocolumn]{article} 
\usepackage{simpleConference}
\usepackage{times}
\usepackage{graphicx}
\usepackage{amssymb}
\usepackage{url,hyperref}

\usepackage{subcaption}
\usepackage{orcidlink}
\usepackage{amsmath}
\usepackage{siunitx}

\newcommand{\mytilde}{\raise.17ex\hbox{$\scriptstyle\mathtt{\sim}$}}

\begin{document}

\hypersetup{hidelinks=true}
\widowpenalty10000

\title{An Evaluation of Multi-Component Weft-Knitted Twill Structures for Sensing Tensile Force}
\author{Roland Aigner\,\orcidlink{0000-0002-8503-4335} \\ Media Interaction Lab \\ 4232 Hagenberg, Austria \\ \href{mailto:roland.aigner@fh-hagenberg.at}{roland.aigner@fh-hagenberg.at} 
   \and Frank Hepper \\ Karl Mayer Stoll GmbH \\ 72770 Reutlingen, Germany \\ \href{mailto:frank.hepper@stoll.com}{frank.hepper@stoll.com} }

\maketitle
\thispagestyle{empty}

\begin{abstract}
We present multi-component knitted resistive sensors for tracking tensile force. The knits were fabricated using a Twill structure, which is a simple pattern featuring anisotropic elastic behavior, providing high stability along course-direction. Our sensors are made of two commercially available conductive yarn types, with highly different linear resistance. We present a variety of integration methods using the proposed Twill structure, all of which can be easily replicated on a two-bed weft-knitting machine. We evaluate the performance of the resulting sensor variations, with respect to consistency, hysteresis, short-term and long-term relaxation and drift, among other metrics. We found that particulars of the knit's loop composition have a crucial effect on the consistency of the sensor readings. Furthermore, we show that knitting resistive yarn more tightly than the substrate material gives superior results and that improving elastic recoil by adding Lycra to the supporting substrate can considerably improve performance. \end{abstract}

\section*{Glossary}

The following is a short and arguably incomplete description of the terms used in the text, however we refrain to go into more detail, since this should be sufficient for the scope of the paper. For more details please refer to \cite{Spencer2001}.

\textbf{knit, tuck, float}: different stitch types performed by the needles. While a \textit{knit} operation forms a new loop by pulling the new yarn through the currently held loop, a \textit{tuck} just adds the yarn to the current loop, i.e., holding/securing the new yarn. In contrast, in a \textit{float} (aka. "miss"), the yarn is guided behind the needle and not held at all.

\textbf{wale, course}: terms describing the dimensions of a knit. Oversimplified but adequate for the scope of this paper, wales and courses can be considered the "columns" and "rows" in a knit, when using matrices as an analogy.

\input{01-intro}
\input{02-sensor}
\input{03-evaluation}

\input{04-results}

\input{05-conclusion}

\section*{Acknowledgment}
This research is part of the COMET project TextileUX (No. 865791, which is funded within the framework of COMET -- Competence Centers for Excellent Technologies by BMVIT, BMDW, and the State of Upper Austria. The COMET program is handled by the FFG.

\bibliographystyle{ieeetr}
\bibliography{main}
\end{document}


\maketitle

\section{Introduction}

This document represents supplementary material for the paper "An Evaluation of Multi-Component Weft-Knitted Twill Structures for Sensing Tensile Force" by Aigner and Hepper \cite{Aigner2023}. The following chapters present more detail on some of the pre-evaluations and methods mentioned in the main paper, which were omitted there to avoid oververbosity.

\section{Pre-evaluations}
\subsection{Correlation between width and height}
\begin{figure}[h]
    \centering
    \begin{subfigure}[T]{.6\textwidth}
        \centering
        \includegraphics[width=1\linewidth]{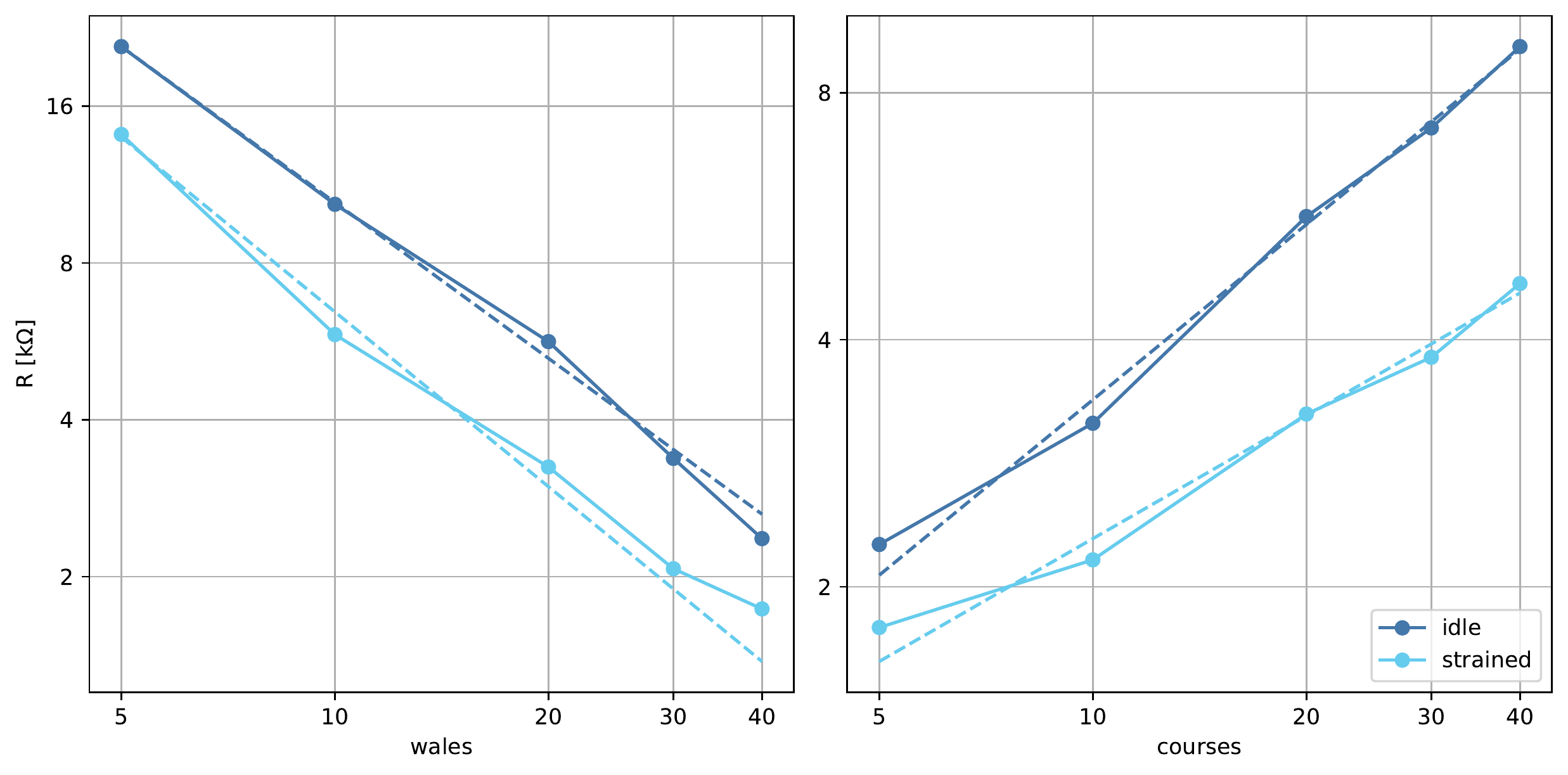}
        \caption{}
        \label{fig:preeval:widthvsheight}
    \end{subfigure}%
    \begin{subfigure}[T]{.33\textwidth}
        \centering
        \includegraphics[width=1\linewidth]{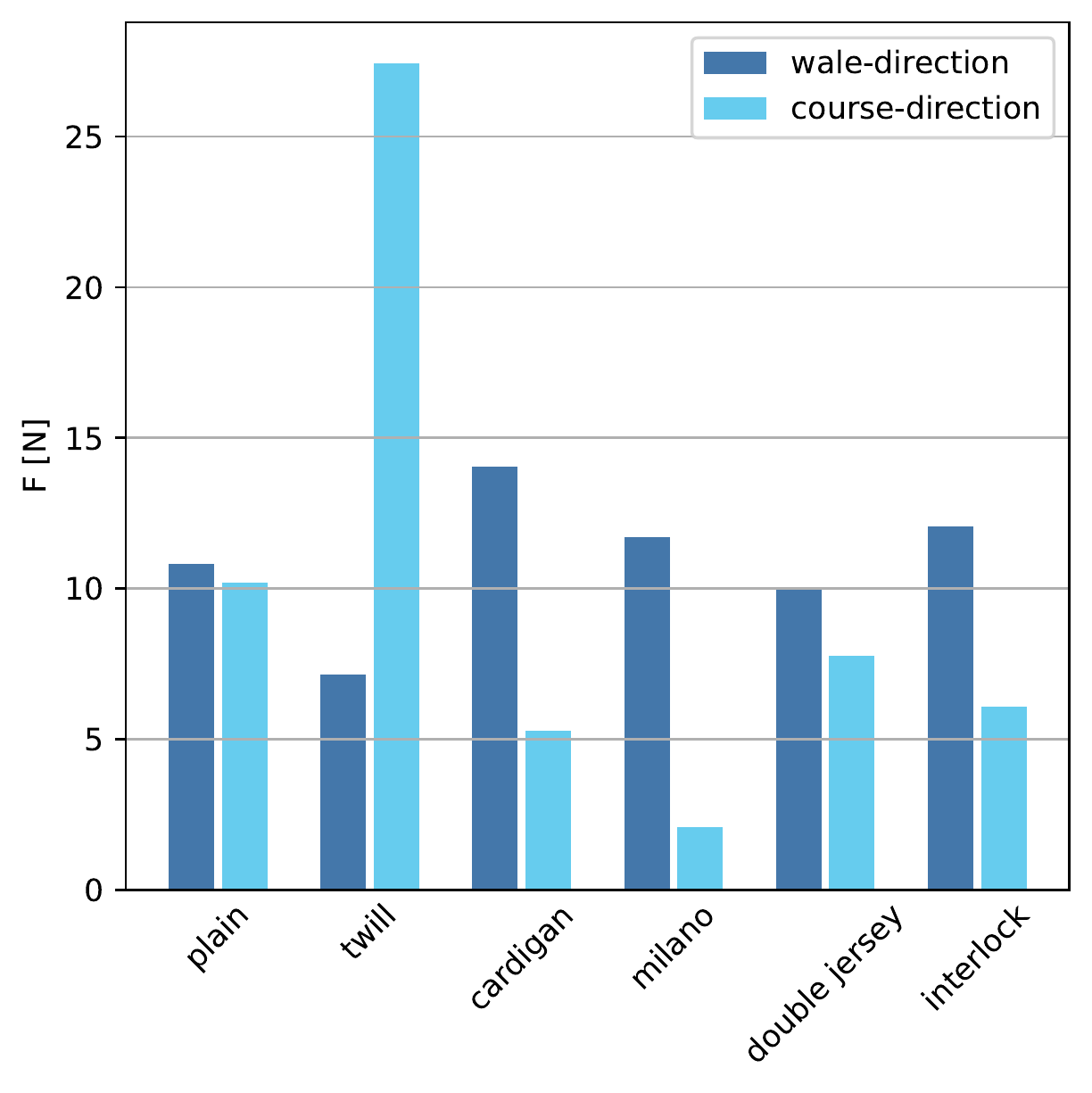}
        \caption{}
        \label{fig:preeval:straintests}
    \end{subfigure}%
    \caption{(a) When the number of sensor wales (left) is changed, the sensor resistance changes proportionally. When the number of sensor courses (right) is changed, the resistance changes inversely proportional. (b) Twill, Cardigan, and Milano show highly anisotropic strain behaviour. As a result of the structural composition, a Twill provides higher stability along course direction, while Cardigan and Milano provide higher stability along wale direction.}
    \label{fig:preeval}
\end{figure}

As mentioned in the paper, in a preliminary evaluation we confirmed the results from \cite{Ou2018}, which found that sensor resistance is directly proportional to height and inversely proportional to width, with
$$
    R=\rho\frac{h}{w}\,,
$$
where $\rho$ is a material-specific constant. Figure \ref{fig:preeval:widthvsheight} illustrates this correlation; we varied the number of wales while keeping the number of courses constant, and vice versa. Resistance values were noted at idle (0\,N) and strained (5\,N) state respectively. Note that resistance values largely deviate from the ones in the main paper, since we knitted different sizes and furthermore used different materials, affecting $\rho$, however knitting structure was identical.

\subsection{Comparison of strain behaviour of different knitting patterns}
We furthermore evaluated strain behaviour in an \textit{ex-ante} evaluation. Using identical stitch settings and material, we knitted patches of 6 common patterns with equal sizes: Plain Knit (aka. Single Jersey), Twill, Cardigan, Milano, Double Jersey, Interlock. We then pulled each of those in our testing apparatus to a strain of e=40\% along wale and along course direction and measured the required force. Figure \ref{fig:preeval:straintests} clearly shows that Twill, Cardigan, and Milano are highly anisotropic regarding straining behaviour. In contrast to Milano and Cardigan, which provide better wale-directional stability due to the frequent tucks, a Twill shows the opposite behaviour, which is a result of the high number of floats.

\subsection{Sensor consistency}

\begin{figure}
    \centering
    \includegraphics[width=1\linewidth]{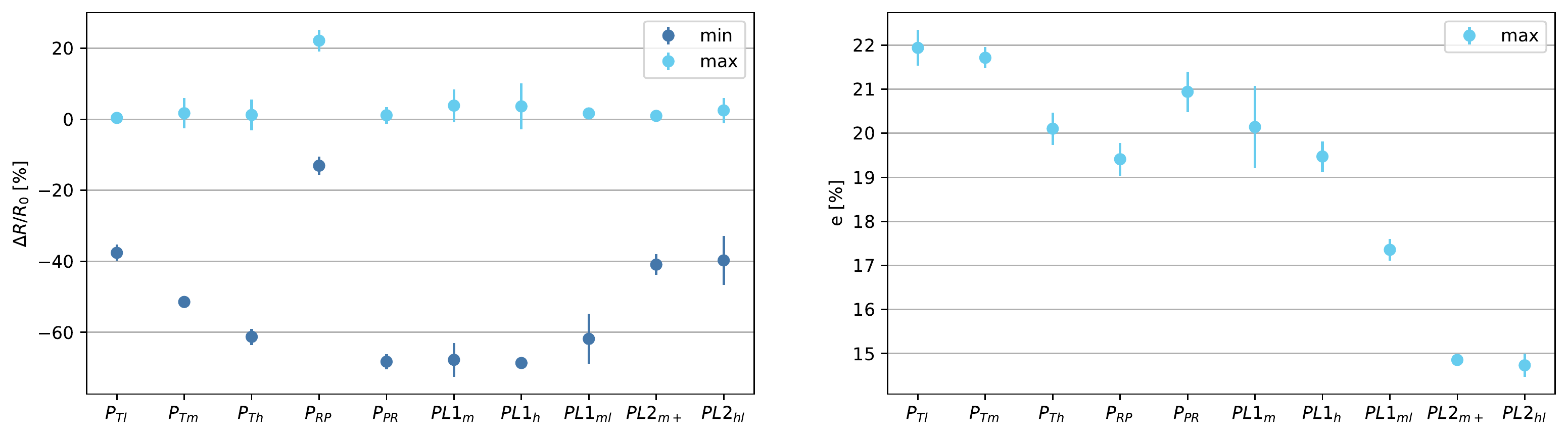}
    \caption{Mean values and SDs for sensor readings (left) and strain (right), when comparing sampling data from multiple sensor patches of the same type show reasonable within-sensor-design consistency.}
    \label{fig:consistency}
\end{figure}

We knitted three samples of each of our sensor designs and therefore briefly investigated within-design consistency. We recorded data from repeated actuation using all of our sensor patches using our testing apparatus, stressing them repeatedly with 10\,N, with our standard jog-rate. We removed the first actuation cycle as an outlier and we compared minima and maxima of $\Delta R/R_0$ out of the remaining data, as well as strain values $e$. Out of our 10$\times$3 patches, we removed 2 outliers (one of type P\textsubscript{Tm} and one of type PL1\textsubscript{ml}) that were showing fabrication related irregularities. Results from the remaining samples are presented in Figure \ref{fig:consistency}, which shows satisfying standard deviations.

\section{Long-term repetition}

\begin{figure}
    \centering
    \includegraphics[width=1\linewidth]{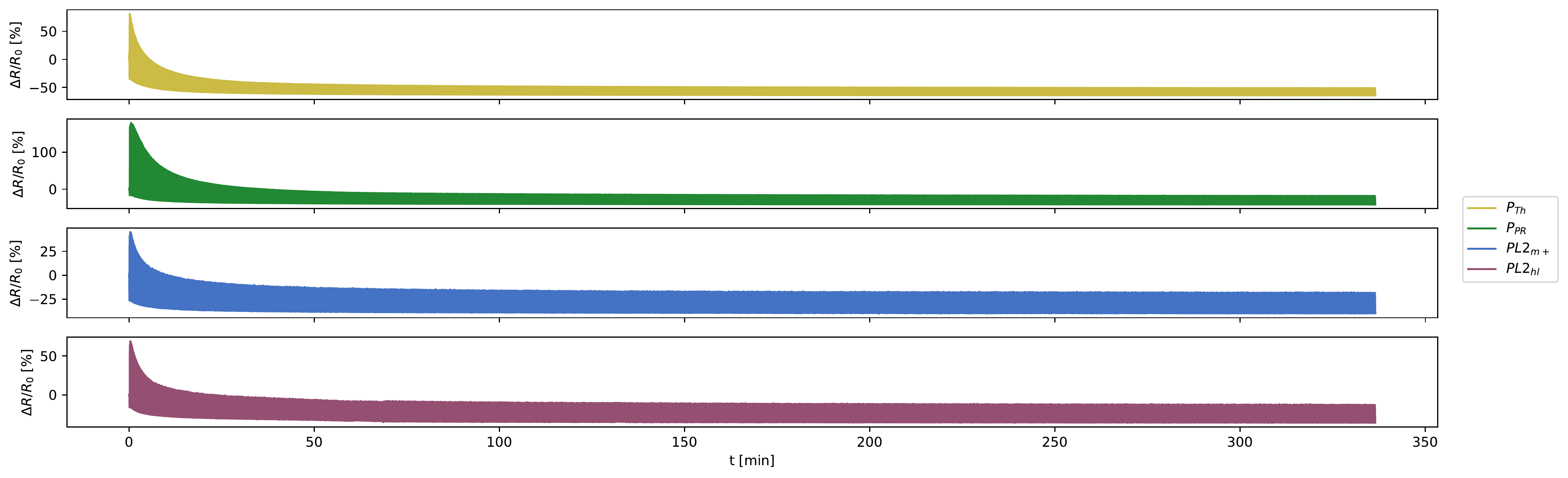}
    \caption{Long-term repeated actuation show that Lycra variants' dynamic range suffer least from constant actuation.}
    \label{fig:longterm-rep}
\end{figure}

Next to long-term drift (see main paper), we tested our best-performing sensors for long-term behaviour at repeated actuation. Using the standard jog rate of our testing apparatus, we repeatedly strained our sensors to e=20\%, dwelling for 3 seconds after each movement (i.e., at e=0\% and e=20\%). We recorded 2,200 actuation cycles, which resulted in a testing duration of approx. 5.6 hours. Results can be seen in Figure \ref{fig:longterm-rep}. The main takeaway is that while settling behaviour is similar, non-Lycra variants' dynamic range suffers most during this procedure. 

\section{Force-related data}
\begin{figure*}
    \centering
    \textit{non-Lycra}\vspace{3pt}\\
    \begin{subfigure}{1\textwidth}
      \centering
      \includegraphics[width=1\textwidth]{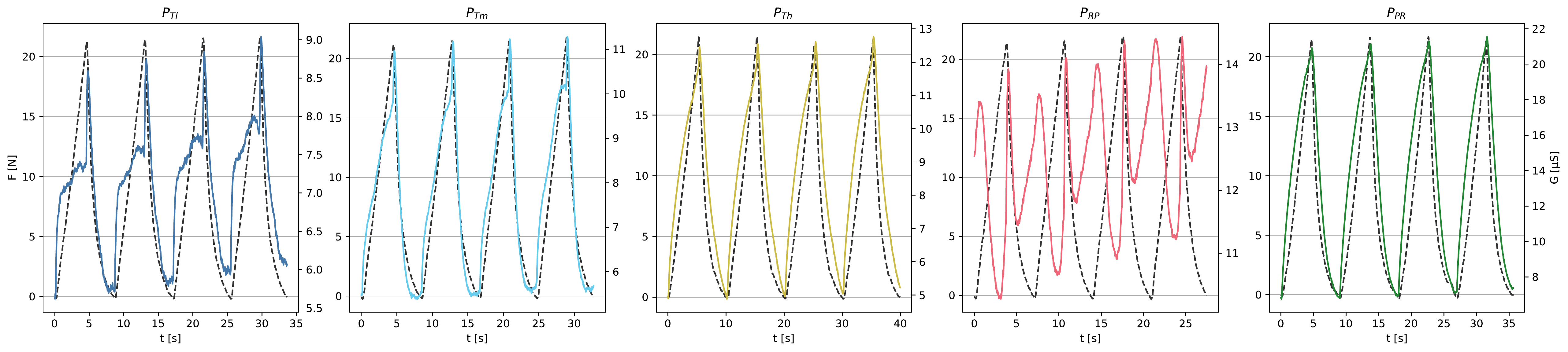}
      \label{fig:timelines:sub1}
    \end{subfigure}
    \textit{Lycra}\vspace{3pt}\\
    \begin{subfigure}{1\textwidth}
      \centering
      \includegraphics[width=1\textwidth]{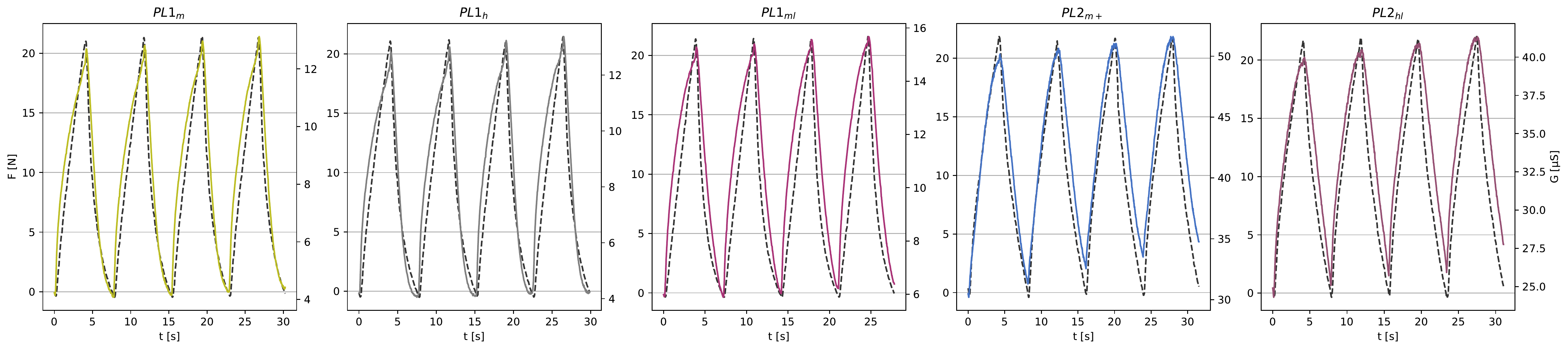}
      \label{fig:timelines:sub2}
    \end{subfigure}
    \textit{course-direction}\vspace{3pt}\\
    \begin{subfigure}{1\textwidth}
      \centering
      \includegraphics[width=1\textwidth]{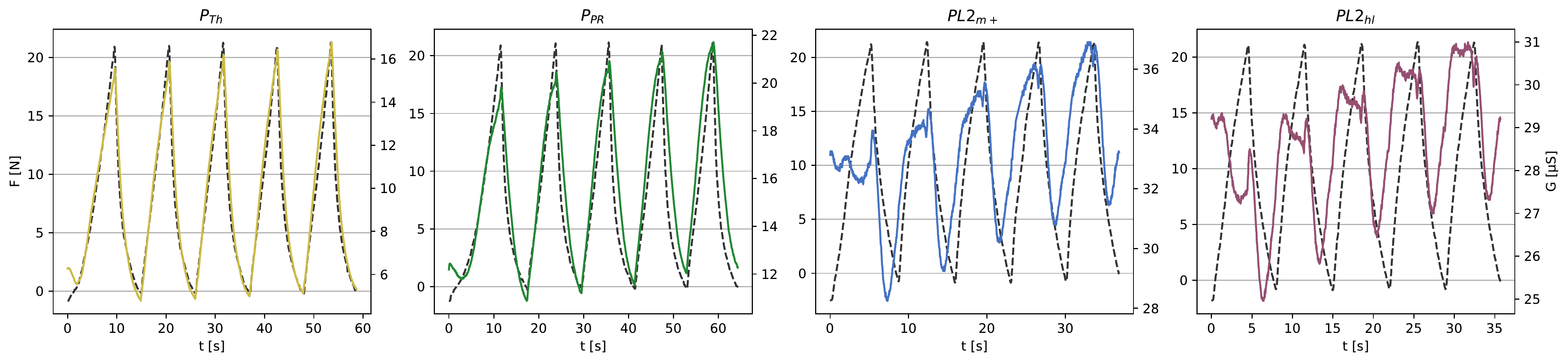}
      \label{fig:timelines:sub3}
    \end{subfigure}%
    
    \caption{Timelines of non-Lycra (top) and Lycra (center) variants: overlaying force $F$ (dashed, black) and sensor conductivity $G$, show respective conformity of our variations. Results of our tests actuating a sub-set along course-direction is presented at the bottom.}
    \label{fig:timelines}
\end{figure*}

\begin{figure*}
  \centering
  \begin{minipage}{0.3\textwidth}
    \centering
    \textit{non-Lycra}
    \includegraphics[width=\linewidth]{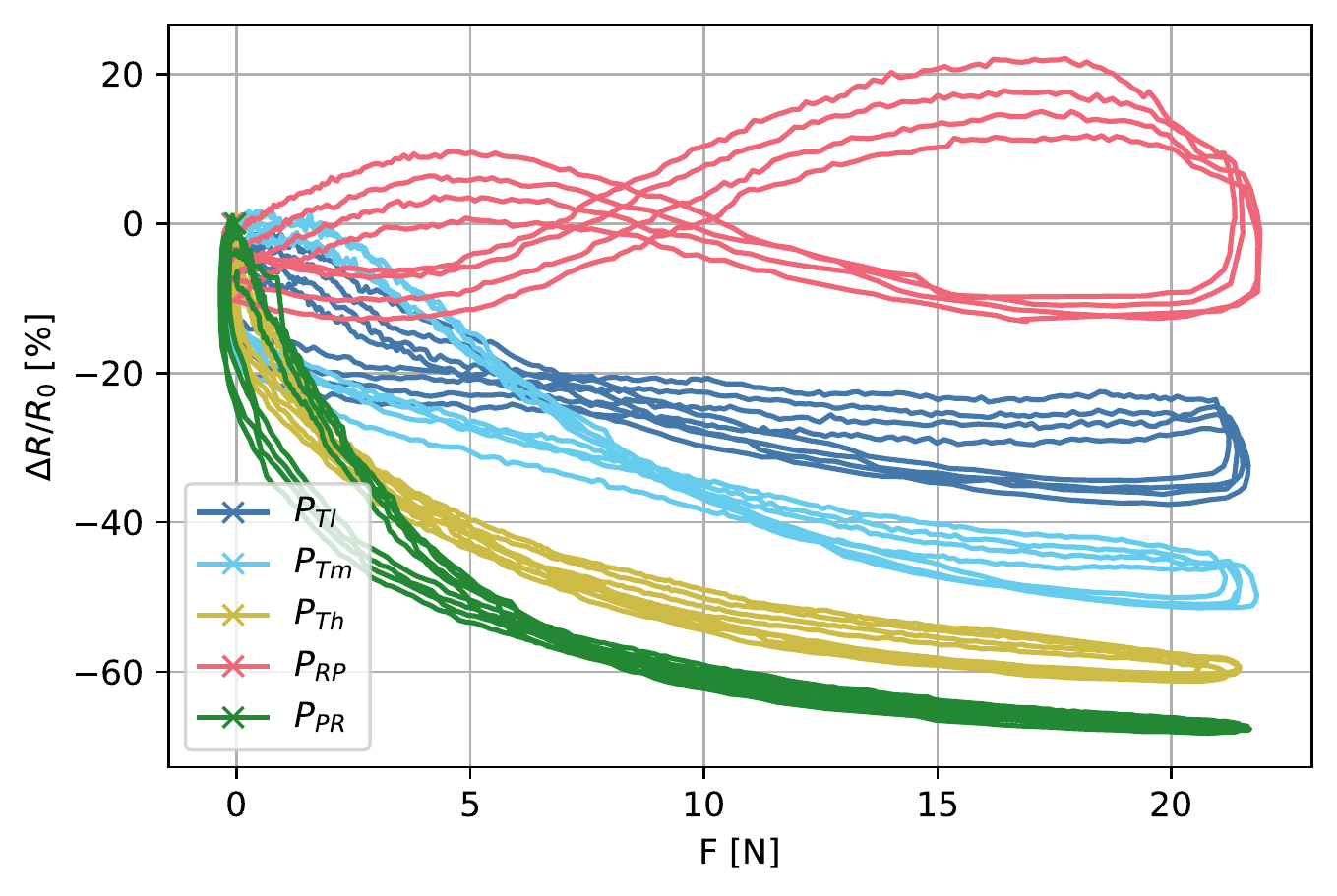}
    \label{fig:characteristics:sub1}
  \end{minipage}%
  \hfill
  \begin{minipage}{0.3\textwidth}
    \centering
    \textit{Lycra}
    \includegraphics[width=\linewidth]{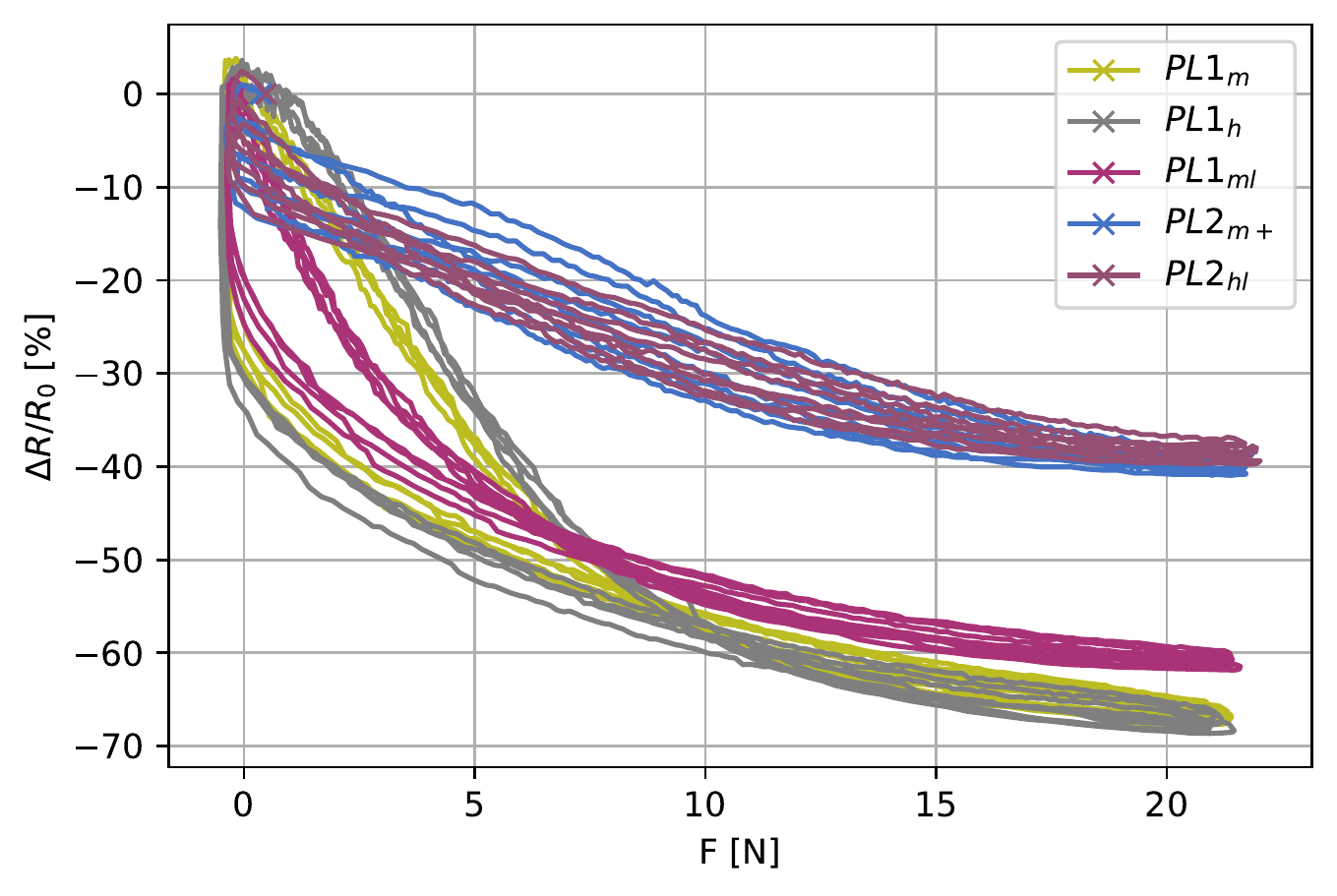}
    \label{fig:characteristics:sub2}
  \end{minipage}%
  \hfill
  \begin{minipage}{0.3\textwidth}
    \centering
    \textit{course-direction}
    \includegraphics[width=\linewidth]{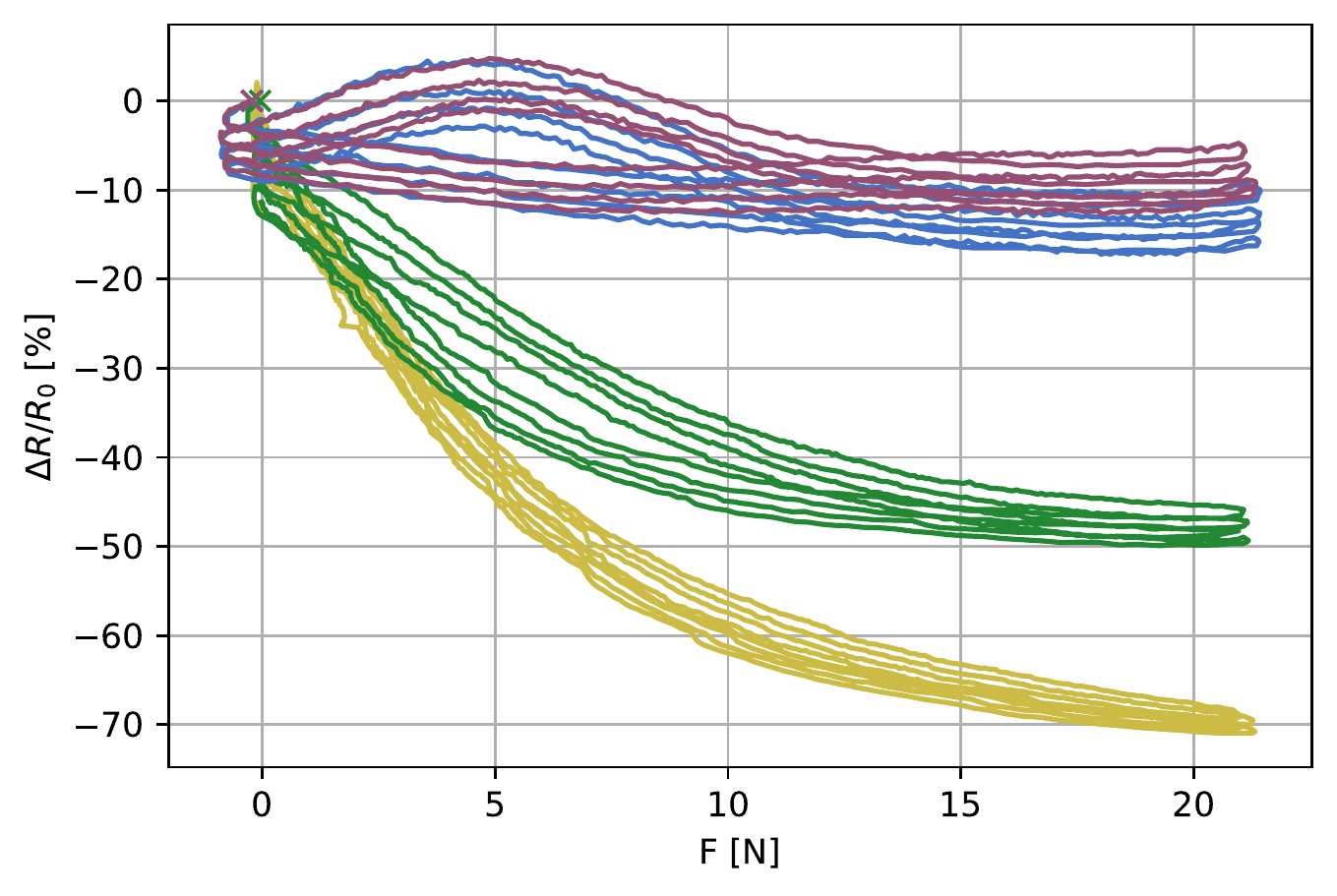}
    \label{fig:characteristics:sub3}
  \end{minipage}
  \caption{Plots of force $F$ against relative resistance change $\Delta R/R_0$, showing characteristics of non-Lycra (left) and Lycra (center) variants, as well as results from course-directional actuation using a sub-set of our variations (right).}
  \label{fig:characteristics}
\end{figure*}

For sake of comparability with related work, we presented plots in the main paper that are mostly related to strain. As mentioned however, our main motivation for this work was to sense \textit{stress}, i.e., force. Hence, in Figure \ref{fig:timelines}, we present timelines with relation to force. Furthermore, we show characteristics in Figure \ref{fig:characteristics}, including result of our course-directional tests of a sensor subset, as described in the main paper.

\section{Details about line-fitting for calculation of hysteresis}

\begin{figure*}
    \centering
    \includegraphics[width=\textwidth]{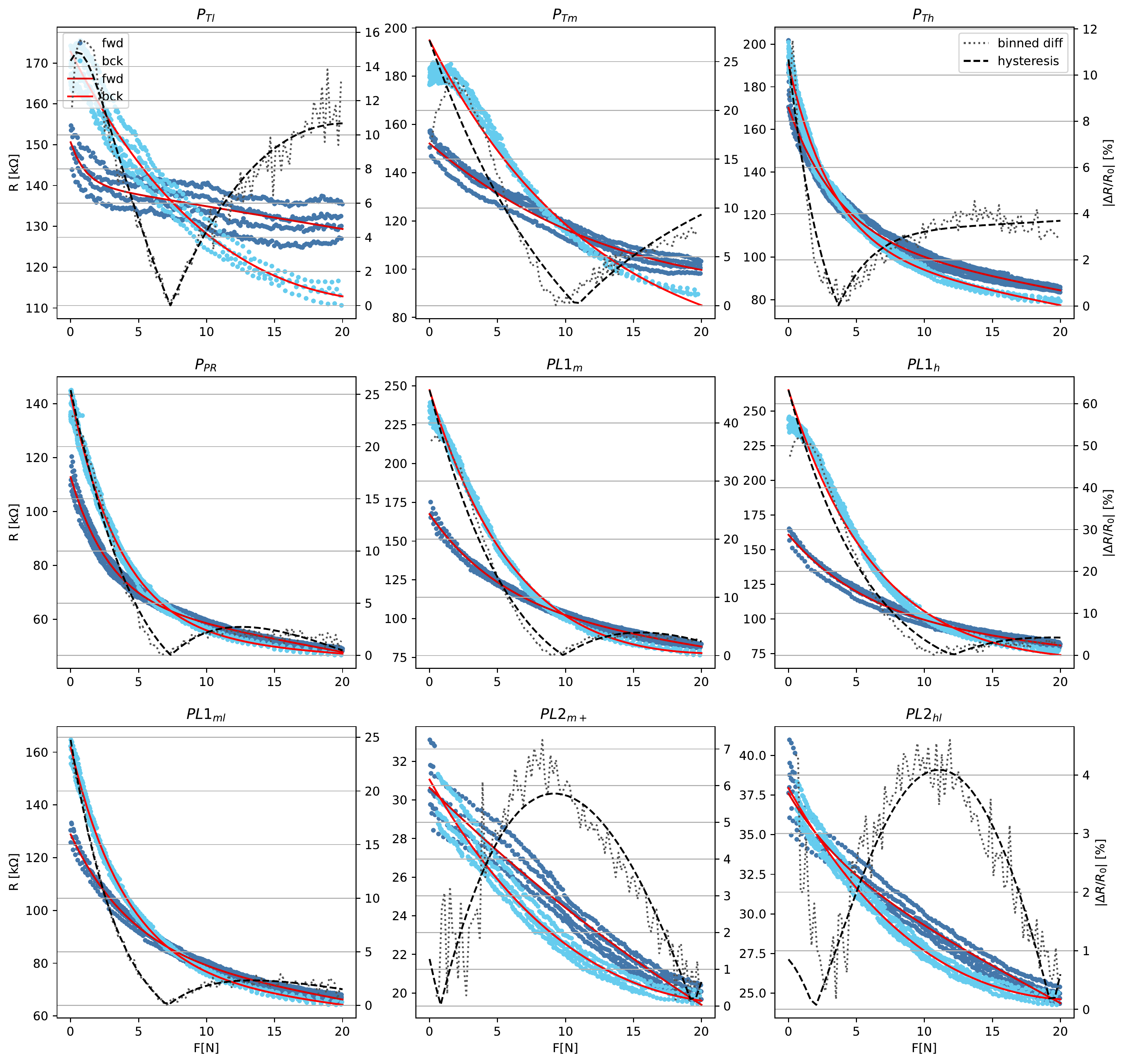}
    \caption{Lines fit to pulling and releasing segments of captured samples for wale-directional ("vertical") testing. Hysteresis in the paper is reported based to the fit curves.}
    \label{fig:hysteresis}
\end{figure*}

\begin{figure*}
    \centering
    \includegraphics[width=\textwidth]{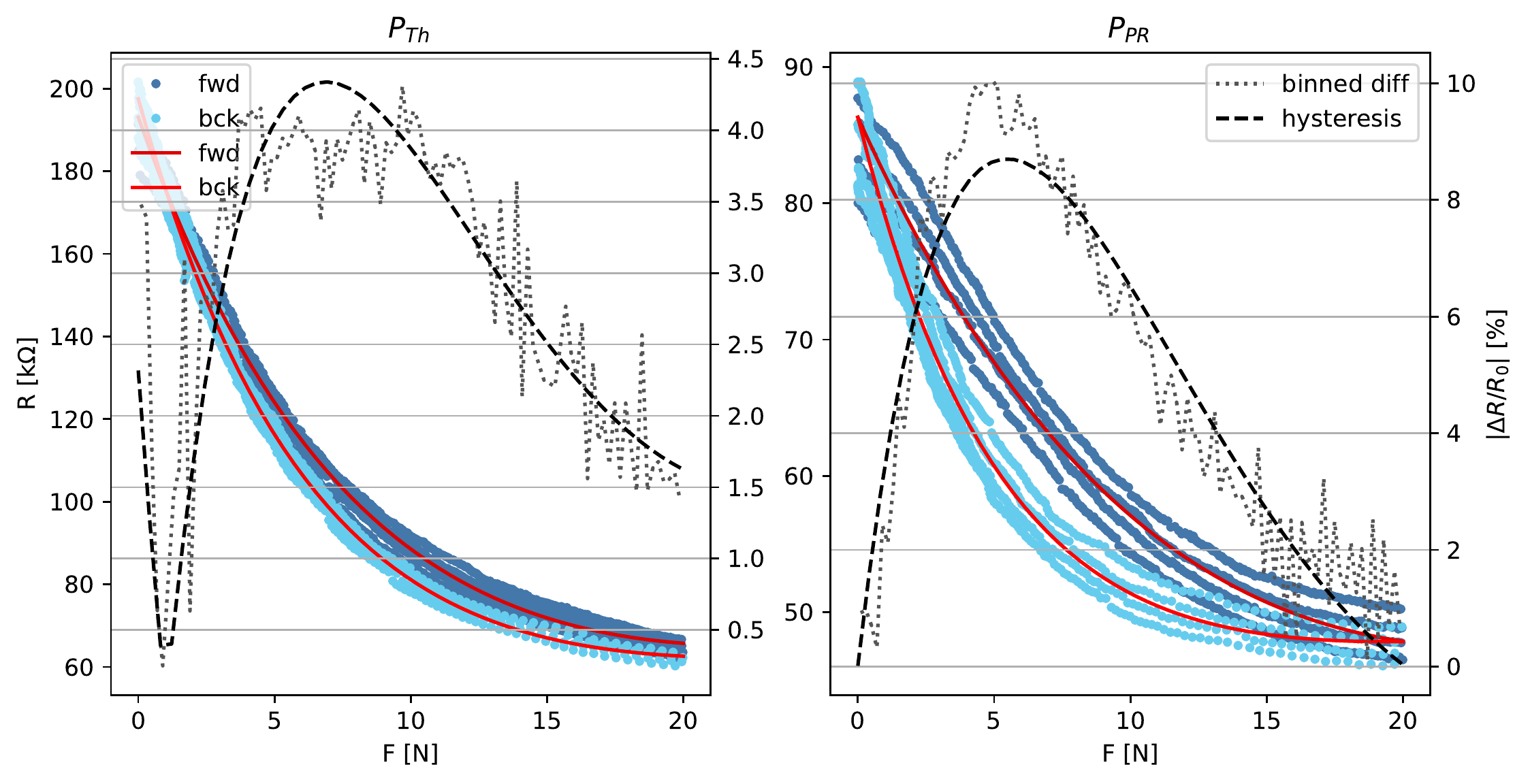}
    \caption{Lines fit to pulling and releasing segments of captured samples for course-directional ("horizontal") testing. Hysteresis in the paper is reported based to the fit curves.}
    \label{fig:hysteresis-course}
\end{figure*}

To perform the line-fitting for calculating max. hysteresis, we first separated data into pulling and releasing segments. We noticed that most trends follow an exponential curve with constant offset, so we tried first to fit functions of type
$$
    y(x) = s \cdot 2^{a(x+o)}+d \,,
$$
using SciPy function \texttt{optimize.curve\string_fit}\footnote{\url{https://docs.scipy.org/doc/scipy/reference/generated/scipy.optimize.curve_fit.html}} however, we noticed that the best fits were still off by a linear function, so we added another linear term $k \cdot x$ to get
$$
    y(x) = s \cdot 2^{a(x+o)}+k \cdot x+d\,.
$$
The resulting fit curves can be seen in Figure \ref{fig:hysteresis} and the respective parameters can be found in Table \ref{tab:fitparams}, along with corresponding $r^2$ values. Since F/R characteristics were highly erratic for P\textsubscript{RP}, we excluded this set. We repeated the same for course-directional pulling, however only for P\textsubscript{Th} and P\textsubscript{PR}, since Lycra-variants PL* were again not performing well (see main paper); the result can be seen in Figure \ref{fig:hysteresis-course}, parameters are found in Table \ref{tab:fitparams-course}.

We calculate hysteresis with the absolute difference between the fit curves. Furthermore, we plot difference of binned data samples. Samples were therefore split the data samples into 100 bins that were equally distributed along the x-axis (F) and calculated mean values for each bin. We did this for pull- and release samples separately and calculated absolute difference, which is plotted as "binned diff" in Figures \ref{fig:hysteresis} and \ref{fig:hysteresis-course}.

\begin{table}
    \centering
    \caption{Parameters of fitted curves for wale-directional testing, as determined using \texttt{curve\string_fit.}}
    \begin{tabular}{|cr||c|c|c|c|c||c|}
        \hline
         & & a & s & d & k & o & r\textsuperscript{2} \\
        \hline
        \hline
        P\textsubscript{tl}
            & pull & -1.35 & 15.8 & 140 & -0.556 & 0.479 & 0.570 \\
            & release & -0.0897 & 100 & 36.3 & 1.86 & -5.00 & 0.952 \\
        \hline
        P\textsubscript{tm}
            & pull & -0.0696 & 100 & 24.8 & 1.32 & -5.00 & 0.958 \\
            & release & -0.112 & 100 & 47.6 & 0.302 & -5.00 & 0.959 \\
        \hline
        P\textsubscript{th}
            & pull & -0.386   & 36.5 & 108      & -1.17   & -2.02 & 0.982 \\
            & release & -0.436   & 43.1 & 102      & -1.26   & -2.36 & 0.988 \\
        \hline
        P\textsubscript{pr}
            & pull & -0.507   & 26.8 & 66.5     & -0.935  & -1.55 & 0.989 \\
            & release & -0.381   & 45.3 & 52.0     & -0.257  & -2.65 & 0.993 \\
        \hline
        \hline
        PL1\textsubscript{m}
            & pull & -0.260   & 38.7 & 103      & -1.14   & -2.81 & 0.991 \\
            & release & -0.210   & 100  & 40.5     &  1.30   & -5.00 & 0.992 \\
        \hline
        PL1\textsubscript{h}
            & pull & -0.233   & 38.8 & 93.8     & -0.778  & -3.38 & 0.982 \\
            & release & -0.217   & 100  & 53.2     &  0.509  & -5.00 & 0.983 \\
        \hline
        PL1\textsubscript{ml}
            & pull & -0.317   & 28.0 & 82.3     & -0.835  & -2.31 & 0.990 \\
            & release & -0.347   & 42.3 & 74.2     & -0.532  & -3.02 & 0.996 \\
        \hline
        PL2\textsubscript{m+}
            & pull & -0.0317  & 27.4 & 2.61e-08 & -0.0178 & -5.00 & 0.956 \\
            & release & -0.0797  & 23.6 & 7.66e-10 &  0.466  & -5.00 & 0.962 \\
        \hline
        PL2\textsubscript{hl}
            & pull & -0.349   & 3.68 & 33.5     & -0.458  & -0.450 & 0.964 \\
            & release & -0.08264 & 28.5 & 2.18e-05 &  0.627  & -5.00  & 0.983 \\
        \hline
    \end{tabular}
    \label{tab:fitparams}
\end{table}

\begin{table}
    \centering
    \caption{Parameters of fitted curves for course-directional testing, as determined using \texttt{curve\string_fit.}}
    \begin{tabular}{|cr||c|c|c|c|c||c|}
        \hline
         & & a & s & d & k & o & r\textsuperscript{2} \\
        \hline
        \hline
        P\textsubscript{th}
            & pull & -0.167 & 97.6 & 19.0  & 1.47 & -5.0 & 0.993 \\
            & release & -0.225 & 77.0 & 40.9  & 0.743   & -4.54 & 0.993 \\
        \hline
        P\textsubscript{pr}
            & pull & -0.0947 & 62.2 & 5.40-15 & 1.23 & -5.0 & 0.968 \\
            & release & -0.271 & 27.5 & 42.6  & 0.212 & -2.46 & 0.969 \\
        \hline
    \end{tabular}
    \label{tab:fitparams-course}
\end{table}

\bibliographystyle{ieeetr}
\bibliography{supplement}

%% file: 01-intro.tex
\section{Introduction}
\label{sec:introduction}


Textile based sensors provide beneficial features such as high flexibility and breathability, which can make them comfortable for wearing them on skin, e.g., when compared to foil-based solutions \cite{Luo2023,Li2022,Chen2020,Shao2019}. This is attractive for use cases requiring long duration of direct skin contact, such as therapy scenarios via bio-monitoring \cite{Atalay2015} and activity tracking \cite{Ayodele2021b,Li2021}, but also for user interfaces such as data gloves, for tracking of hand posture \cite{Akerfeldt2015,Han2019} or gesture detection \cite{Lee2021}. Many of those use cases are already implemented with knitted fabrics, since they are particularly suitable for sensing strain due to their innate stretchability.

\begin{figure}[!t]
    \centering
    \begin{subfigure}{.66\columnwidth}
      \centering
      \includegraphics[width=0.95\linewidth]{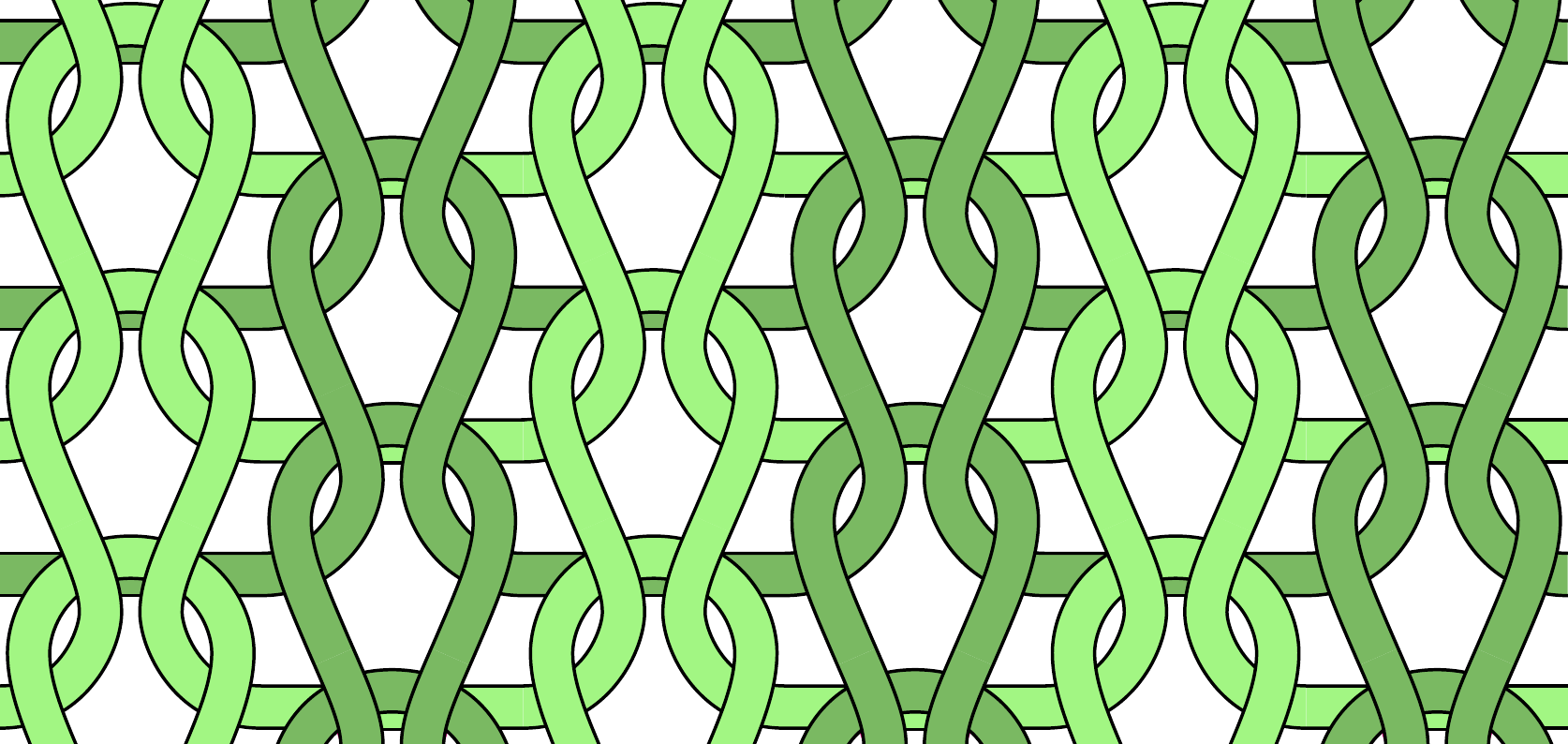}
      \caption{}
      \label{fig:knit-circuit:sub1}
    \end{subfigure}%
    \begin{subfigure}{.33\columnwidth}
      \centering
      \includegraphics[width=0.95\linewidth]{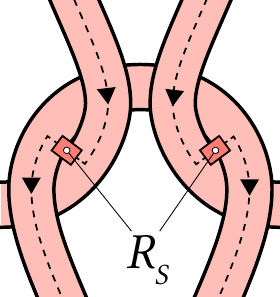}
      \caption{}
      \label{fig:knit-circuit:sub2}
    \end{subfigure}
    \begin{subfigure}{1\columnwidth}
      \centering
      \includegraphics[width=0.95\linewidth]{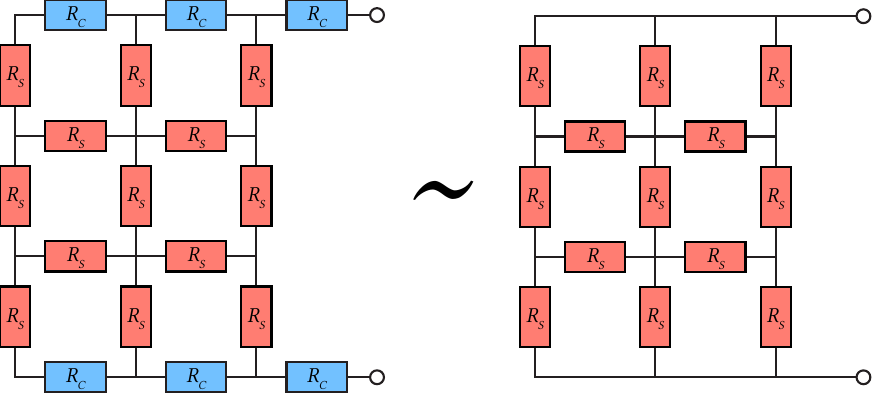}
      \caption{}
      \label{fig:knit-circuit:sub3}
    \end{subfigure}
    \caption{Illustration of a Twill knitting pattern (a). Where current flows along the yarn, loop intermeshing points act as variable resistors (b), increasing conductivity with physical stress at the contact positions. The overall knit geometry can therefore be modeled as a network of variable resistors (c). We maximize the ratio of sensor yarn resistivity and connector yarn resistivity, so changes in connecting parts' resistance $R_C$ are negligible over the much higher absolute values from sensor loops $R_S$.}
    \label{fig:knit-circuit}
\end{figure}

Although there already is a large body of work focusing on knitted strain sensors, we found that most of them are based on highly stretchable fabrics, which are not always desirable. Many scenarios, also beyond garments, require solutions that provide higher tensile stability, or even anisotropic elasticity. Those properties are mostly subject to the geometric composition of the knit, i.e, the knitting pattern. In weft-knitting, those pattern can be thoroughly engineered down to loop level, which is in contrast to warp-knitting \cite{Grassi2017}. Examples for patterns with relatively high extensibility are Plain Knit (aka. Single Jersey) \cite{Liang2019,Lin2015,Tohidi2018,Wicaksono2022,Zhang2006}, Double Jersey \cite{Liang2019}, and rib structures \cite{Bozali2021,Liang2019,Yang2009}. In contrast, patterns with higher stability are relatively rare in related literature, examples for those are Interlock used by Atalay et al. \cite{Atalay2015}, and Cardigan used by Ehrmann et al. \cite{Ehrmann2014}, which showed they provide better sensitivity in low-elongation ranges, when compared to Double Jersey. However, both Interlock and Cardigan represent patterns that occupy both of the machine's needle beds at all times, which is a potential limitation in flexibility for fabrication and design. Therefore, and in contrast to the stated works, we use a pattern that is inconsistently called \textit{Twill} in the textile industry, due to its structural similarity with the weave pattern of the same name. It is a generally widespread and simple pattern, however not touched in textile sensor literature, to our knowledge. As illustrated in Figure~\ref{fig:knit-circuit:sub1}, it consists of courses of alternating knit and float stitches, while the sequence is shifted by one needle for every other course. Due to this high number floats, it provides exceptional stability in course-direction (i.e., "horizontally"), while being more extensible along wale-direction (i.e., "vertically"), when compared to Plain or Double Jersey Knits. This represents a distinct property of a Twill, as opposed to Cardigan or Interlock, with exactly opposite behavior, as a preliminary study confirmed (see supplement). Note that manufacturing of a Twill only requires one needle bed, which increases flexibility and can be a design advantage over other patterns. For example, using a two-bed machine, the entirety of the sensing part can be hidden away to one face of the knit, as done in the work presented in this paper, which can be of aesthetic preference and/or protect it from exposure and therefore from abrasion and damage.

In contrast to most of related work, which focuses on sensing strain \cite{Zhang2006,Akerfeldt2015,Liang2019,Chen2019,Ayodele2021c,Bozali2021,Liang2022}, our primary interest is in sensing \textit{force}, which cannot be trivially inferred from strain directly, due to short-term wear-out effects, exposing hysteresis. Furthermore, knits are subject to considerable structure-dependent relaxation, causing a gradual, non-linear decrease of force at constant elongation. Based on our observations during this work, we noticed that recorded displacement values (and thus inferred strain values) are not entirely adequate to reflect the state of the fabric, since it may be slack when the actuator returns to its initial position. We argue that due to these effects, it would be necessary to record and reconstruct the true fabric lengths by different means, e.g., by optically tracking its geometric state, which would complicate not only the setup, but also reporting and its comprehensibility. Within our work, we consequently investigate the sensors' response to force directly and thus avoid this issue. For the sake of comparibility with related work however, we still include strain data in this paper.

By combining two types of conductive yarn that are knitted directly into the fabric, we produce a fully functional textile force sensor without requiring manual finishing steps. This is opposed to augmenting pre-existing knits by embroidering \cite{Akerfeldt2015} or printing \cite{Jang2021} functional parts, or by sewing patches of conductive fabric\cite{Teyeme2020}. Other works demonstrate the method of polymerizing parts of textiles with Polypyrrole, e.g., the seminal work of DeRossi et al. \cite{DeRossi2002} showed a data glove with resistive sensing areas. However, this process is challenging to do in a consistent manner, when compared to computerized flatbed knitting, which provides loop-level control. Hence, in contrast to the stated works, our method enables to precisely design and tune the sensor structure even to create highly intricate sensor shapes and complex connector traces (cf. \cite{Semjonova2022}).

The goal of our work was to find the optimal variant with respect to general sensor consistency (for repeated actuation with equal or varying force, as well as different actuation speeds), hysteresis, dynamic range, offset, relaxation, drift, and anisotropic behavior. We therefore explored different implementations and contrasted their behavior in a systematic evaluation.

In a nutshell, the top contributions of this paper are:
\begin{itemize}
    \item Three methods of integrating a Twill-based resistive sensor on a Twill substrate fabric, including conductive knit connector traces for attaching electronics at remote positions.
    \item 10 variations of these sensor designs, using different substrate material compositions and yarn tensions.
    \item An in-depth evaluation of those 10 variations and our findings regarding consistency, relaxation, offset, and drift in different scenarios.
\end{itemize}

\begin{figure}
    \centering
    \includegraphics[width=1\linewidth]{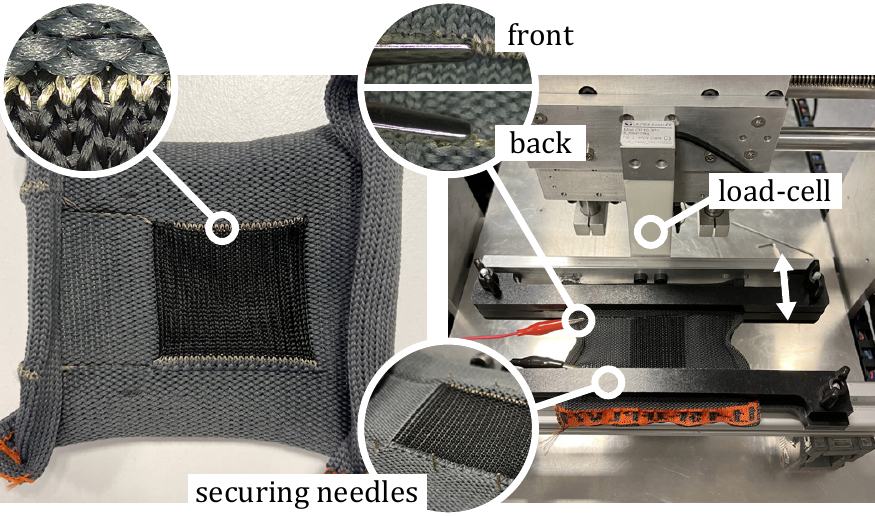}
    \caption{Sample of one of our sensor patches (left), with conductive yarn traces connecting the resistive area (black) on both upper and lower ends. We evaluated our sensors using a custom-built tensile tester with an integrated force cell (right).}
    \label{fig:sensor-tester}
\end{figure}

%% file: 02-sensor.tex
\section{Sensor Implementation}
\label{sec:sensor}

\begin{figure*}[!t]
    \centering
    \begin{subfigure}{.33\textwidth}
      \centering
      \includegraphics[width=1\textwidth]{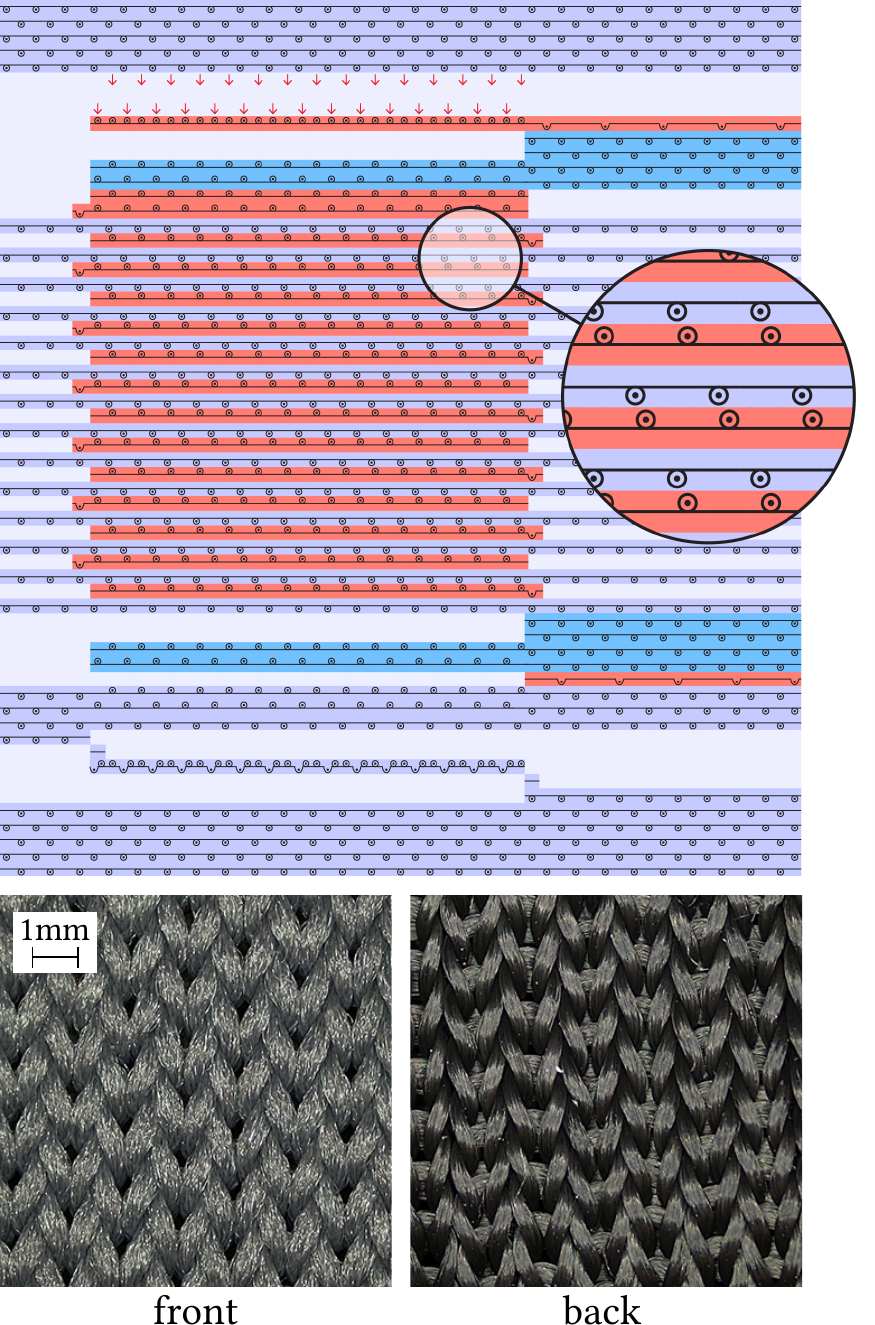}
      \caption{}
      \label{fig:patterns:sub1}
    \end{subfigure}%
    \begin{subfigure}{.33\textwidth}
      \centering
      \includegraphics[width=1\textwidth]{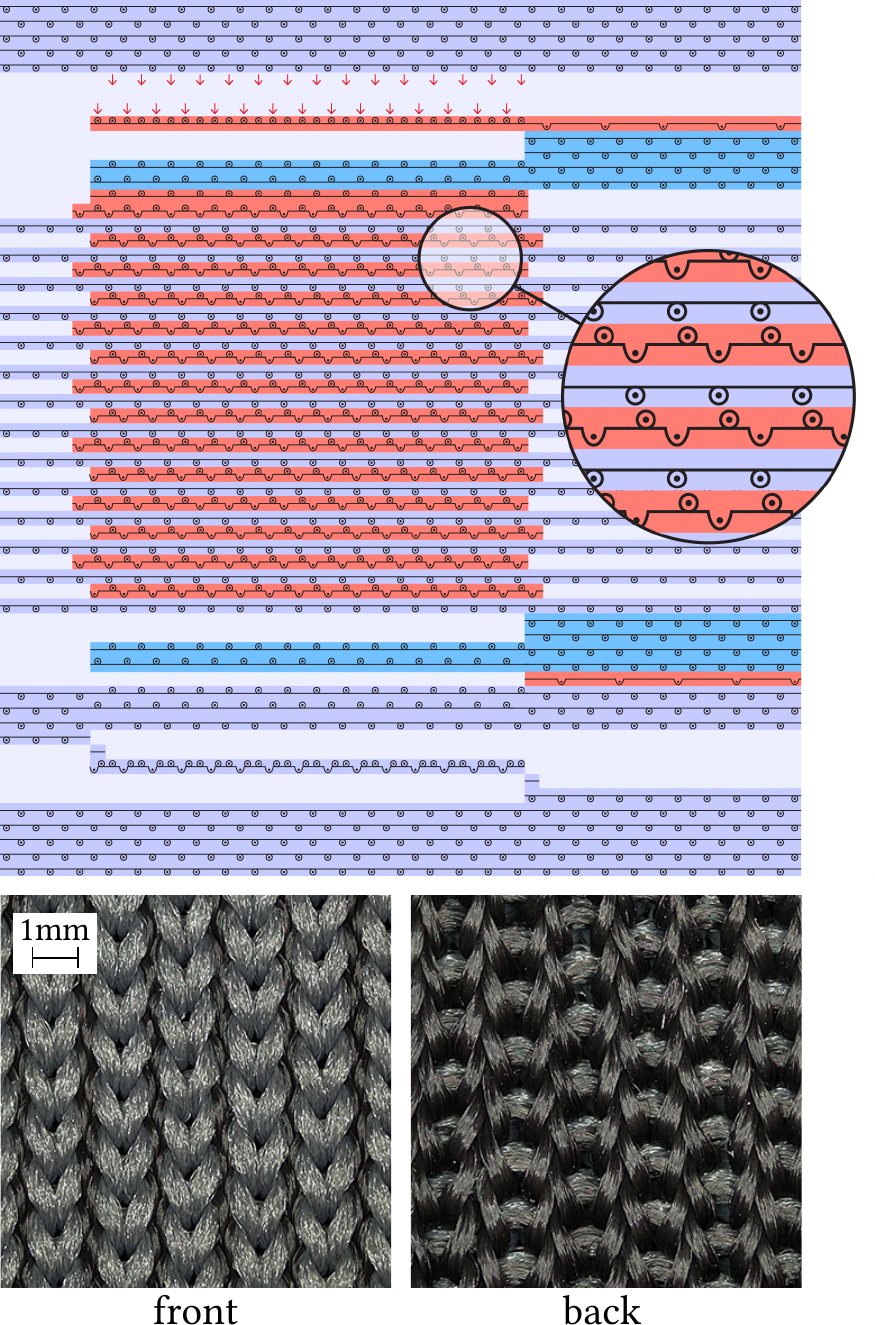}
      \caption{}
      \label{fig:patterns:sub2}
    \end{subfigure}%
    \begin{subfigure}{.33\textwidth}
      \centering
      \includegraphics[width=1\textwidth]{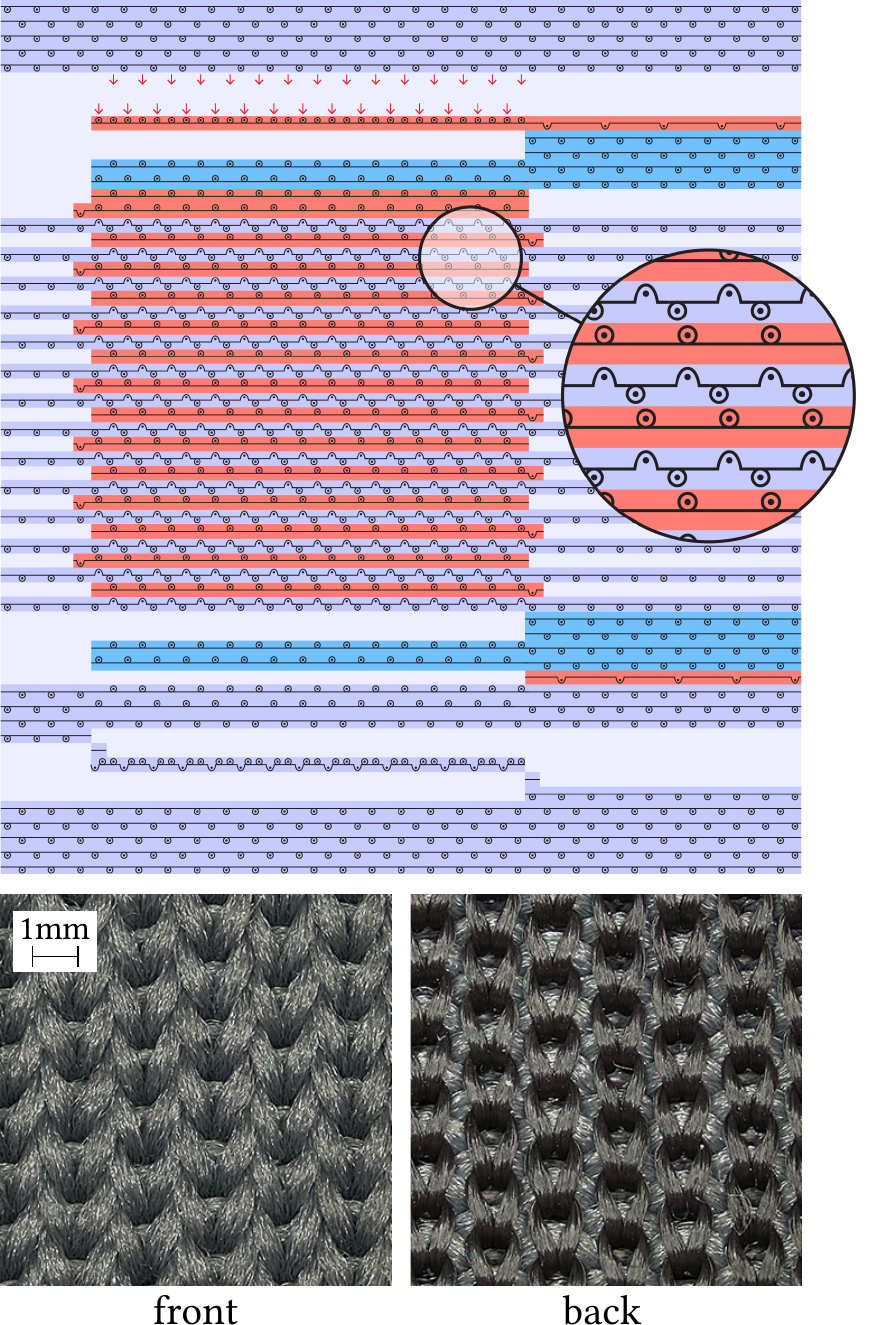}
      \caption{}
      \label{fig:patterns:sub3}
    \end{subfigure}
    \caption{Twill based knitting structures for T (tubular, a), P$\leftarrow$R (Resistat tucked to front-face PES, b), and P$\rightarrow$R (PES tucked to back-face Resistat, c). Note the connecting front bed tucks at the beginning of each Resistat row (red) which secures the edges of the sensor area with the substrate knit (purple). Connector traces (blue) are knit on the back bed for connecting to the Resistat loops, and on the front bed otherwise. Images at the bottom show closeups of front bed (PES) and back bed (Resistat) faces of the resulting knit structures. In particular, the closeups show P\textsubscript{Tm} (a), P\textsubscript{RP} (b), and P\textsubscript{PR} (c).}
    \label{fig:patterns}
\end{figure*}

In the following chapter, we outline the sensor design, including potential for slight modifications that we expected to have an impact on sensor performance. We present our knitted samples and specify manufacturing details. All our patches where knit on a flat-bed knitting machine of type ADF 530-32 KI W Multi Gauge from KARL MAYER STOLL, at gauge E\,7.2. Knitting programs were created with Patternsoftware M1 PLUS Version 7.5.

\subsection{General Sensor Design and Sensing Principle}

The functional principle of our knit force sensors is according to Holm's Theory \cite{Holm1967}, which states that contact resistance is depending on material resistivity $\rho$ and hardness $H$, as well as contact point count $n$ and pressure $P$, with
    $$R=\frac{\rho}{2}\sqrt{\frac{\pi H}{n P}} \,.$$
Since contact pressure between loops is varying (and may even be zero when loops lose contact), the resistance drops when force is applied. Consequently, each intermeshing point in the sensor knit can be considered a variable resistor (cf. Figure~\ref{fig:knit-circuit:sub2}), and moreover, the overall structure can be modeled as a network of resistors, as done for analytical solutions by \cite{Li2010,Liu2016,Ayodele2021c}. 

Similar to Baribina et al. \cite{Baribina2018} and Semjonova et al. \cite{Semjonova2022}, we utilize a multi-material sensor layout, combining conductive and resistive yarn, i.e., two types with largely different linear resistance, which provides several advantages:

\begin{itemize}
    \item Conductive parts can be utilized to knit connector traces, that enable to comfortably attach readout electronics at remote positions of the fabric. This is unlike other work that requires attaching of connecting wires directly at the sensing structures, such as \cite{Lee2021,Li2021}. 
    \item In increasing the resistance ratio of sensor area to connector trace we ensure that the sensor area operates in vastly different absolute value ranges, when compared to the connecting parts. Hence, resistance changes caused by deformation of the connectors are minor and therefore negligible, when compared to the sensor's operational range. This is similar to \cite{Baribina2018} and addresses an issue often ignored in related work, while furthermore enabling to create more explicit and localized force sensors on a textile.
    \item Connecting the resistive part with conductive yarn along the entire width (cf. Figure~\ref{fig:sensor-tester} left) provides more uniform current flow across all wales (cf. Figure~\ref{fig:knit-circuit:sub3}), since R\textsubscript{C} is insignificant against R\textsubscript{S}. This should improve sensitivity consistency across the whole sensor area, which could be particularly relevant when the number of sensor wales is much higher than the number of its courses.
\end{itemize}

\subsection{Materials}

As a resistive yarn for the sensor areas, we used Shakespeare\textsuperscript{\textregistered} Resistat P6204 H100i\footnote{\url{https://shakespeare-pf.com/product/polyester/}}, which is a den\,100/24 Polyester fiber with Carbon sheath, providing relatively high linear electrical resistance of \mytilde 10\,M$\Omega$/m. We twisted four den\,100/24 threads with 30S in 1\textsuperscript{st} stage and 50Z in 2\textsuperscript{nd} stage, to achieve adequate yarn count for a balanced knit when combined with our PES yarn, yielding a den\,400 thread with den\,100/24x4 and \mytilde 2.5\,M$\Omega$/m.

The conductive traces for providing connections were knit with Shieldex\textsuperscript{\textregistered} Madeira HC40\footnote{\url{https://www.shieldex.de/products/madeira-hc-40/}}, which is a silver-coated PA yarn with den\,260 and electrical resistance of $<$300\,$\Omega$/m and proved highly durable during previous work \cite{Aigner2022,Pointner2022}.

For the surrounding substrate base structure, we used a PES with den\,150/30 (TWD Fibres GmbH). The Lycra that we plated along the PES for improving the fabric's elastic recoil was a den\,140 Lycra core covered with PES den\,150/20 (Jörg Lederer GmbH).


\subsection{Knit Structure and Manufacturing}

As any pattern that is knit on a single bed, internal forces on the Twill are unbalanced, meaning it shows inherent curling tendency. For many use case scenarios, where the knit is tailored together with other parts, this may not be an issue. Otherwise, it can be counteracted by framing with a more stable knit structure. 

Instead of implementing the sensor area as an Intarsia field within a surrounding PES structure (cf. \cite{Li2010,Lin2015,Han2019,Isaia2020,Lee2021}), we knit the resistive yarn on the opposite needle bed and connect it to the PES face, which is knit as a continuous Twill. Apart from a more straight-forward integration into a knit, this provides better control about the force distribution throughout the structure. Due to different properties of PES and Resistat, an Intarsia field requires proper tuning of yarn count and stitch settings, to prevent an unbalanced and non-uniform surface. By knitting two faces on opposite beds, we gain more flexibility in tuning the Resistat tightness -- and therefore the sensor's responsiveness -- without introducing areas of considerable physical, visual, and haptic inconsistency. A side-effect is also that the functional parts can be hidden away and are therefore protected from abrasion by the covering PES layer, which may be a benefit in some use cases. 

For connecting the sensor face with the base structure, we tuck the Resistat to the opposite bed at the beginning of each knit course. On upper and lower courses, the Resistat is knit to the conductive yarn which provide connector traces. Figure\,\ref{fig:patterns} provides a detailed knitting diagram. For knitting, we plied 2 threads of den\,400 Resistat for the resistive parts, 2 threads of den\,260 Madeira HC40 for the conductive parts, and 6 threads of den\,150 PES for the substrate.

\begin{table*}[t]
    \centering
    \caption{Overview of our sensor variations. We varied structure composition (Type), ratio of PES threads vs. Lycra threads, nominal stitch length (NP) of substrate material (PES+Lycra), Resistat material (Res), and tuck stitches that connect front and back faces. Note that NP are a measure of yarn usage per loop, i.e., lower numbers represent tighter knits.}
    \begin{tabular}{|l|c|c|c|c|c|c|l|}
        \hline
        Name & Type$^\ast$ & PES & Lycra & NP PES & NP Res & NP tuck & notes \\
        \hline
        P\textsubscript{Tl} 
            & T
            & 6  & -     
            & 13.1 & 13.5 & -
            & tubular, low tension for Resistat \\
        P\textsubscript{Tm} 
            & T   
            & 6  & -
            & 13.1 & 12.5 & -
            & tubular, medium tension for PES/Resistat \\
        P\textsubscript{Th} 
            & T   
            & 6  & -     
            & 13.1 & 11.5 & - 
            & tubular, high tension for Resistat \\
        P\textsubscript{RP} 
            & P$\leftarrow$R 
            & 6  & -    
            & 13.1 & 12.0 & 9.0 
            & Resistat tucked to PES \\
        P\textsubscript{PR}
            & P$\rightarrow$R 
            & 6  & -     
            & 13.1 & 12.0 & 9.0
            & PES tucked to Resistat \\
        \hline
        PL1\textsubscript{m}
            & P$\rightarrow$R 
            & 5  & 1 
            & 12.5 & 12.5 & 9.0 
            & 1$\times$Lycra + medium tension Resistat \\
        PL1\textsubscript{h}
            & P$\rightarrow$R 
            & 5  & 1
            & 12.5 & 11.5 & 9.0
            & 1$\times$Lycra + high tension Resistat \\
        PL1\textsubscript{ml}
            & P$\rightarrow$R 
            & 5  & 1
            & 12.5 & 12.5 & 9.5 
            & 1$\times$Lycra + medium tension Resistat, low tension tuck \\
        PL2\textsubscript{m+}
            & P$\rightarrow$R 
            & 4  & 2
            & 12.0 & 11.8 & 9.0   
            & 2$\times$Lycra + medium-high tension Resistat \\
        PL2\textsubscript{hl}
            & P$\rightarrow$R 
            & 4  & 2
            & 12.0 & 11.5 & 9.5 
            & 2$\times$Lycra + high tension Resistat, low tension tuck \\
        \hline
        \multicolumn{8}{l}{$^\ast$ Types: T = tubular knit; P$\leftarrow$R = Resistat is tucked to front-face PES; P$\rightarrow$R = PES is tucked to back-face Resistat}
    \end{tabular}
    \label{tab:knitVariations}
\end{table*}

\subsection{Variations}
\label{subsec:variations}

As mentioned above, we knitted non-functional PES and functional Resistat on opposite needle beds, yielding two knit faces that need to be fixated so they do not fall apart. We investigated three options of doing so: the most straight-forward one is to connect both faces along the sensor's outer wales (cf. Figure \ref{fig:patterns:sub1}), by tucking at the respective outer needles. This results in a "tubular" knit structure, which we henceforth will address with "T". Note that both faces are completely detached in this knit, which could lead to erratic behaviour, depending on the fabric's firmness. We therefore tried three variations with the sensor parts knitted with different tightness: one with the Resistat knit with \textit{lower} tension than the PES ("Tl"), one with \textit{medium} tension, meaning PES and Resistat tightness balanced ("Tm"), and one with Resistat knit with \textit{higher} tension than PES ("Th"). 

We furthermore created variations that kept both faces closely attached, by tucking the Resistat to the PES ("P$\leftarrow$R") across the entire courses (cf. Figure \ref{fig:patterns:sub2}), as well as the opposite, tucking the PES to the Resistat loops ("P$\rightarrow$R", cf. Figure \ref{fig:patterns:sub3}). 

From handling with the resulting knits, we could subjectively see that our first variations with 6 threads of PES ("P") were prone to short-term wear-out and we therefore expected poor elastic recoil. For this reason, we also created samples that combined PES with Lycra ("PL"), to encounter this aspect (cf. Table \ref{tab:knitVariations}). Hence, in addition to using 6$\times$PES for the surrounding substrate, we also created patches where we plated 5$\times$PES together with 1$\times$Lycra ("PL1"), as well as 4$\times$PES with 2$\times$Lycra ("PL2"). We already saw during our first evaluation, that lower-tension Resistat patches performed poorly, as well as connecting front and back faces outperforms tubular structures, we therefore chose to focus on those for our Lycra variations, hence, all of them were knit of type P$\rightarrow$R with \textit{medium} to \textit{high} Resistat tension.

%% file: 03-evaluation.tex
\section{Evaluation}
\label{sec:evaluation}

\subsection{Apparatus}

For evaluation, we used a custom tensile tester, which we built from an obsolete CNC milling machine (cf. Figure\,\ref{fig:sensor-tester} right). The machine was operated by Art-Soft Mach4 CNC Control Software (v4.2.0), running on a Windows\ 10 PC. We attached mounts to clamp the textiles on both ends, incorporated needles at 2\,cm distance additionally secured the textile so it would not slip. The clamp attached to the moving part was equipped with a single-point load-cell of type Sauter CP P1-Ba-d-1810\footnote{\url{https://www.kern-sohn.com/shop/en/measuring-instruments/measuring-cells/CP-P1/}} which was sampled at \mytilde 80\,Hz with an ADS\ 1231 24-bit Delta-Sigma ADC\footnote{\url{https://www.ti.com/product/ADS1231}}. We acquired the sensors' resistance values using a simple voltage divider with a reference resistor of 606\,k$\Omega$ and sampled using an Adafruit ADS1115 16-bit ADC\footnote{\url{https://www.adafruit.com/product/1085}} at \mytilde 128\,Hz. At \mytilde 40\,Hz, we averaged the samples of the previous period, and captured the results into CSV files for later analysis, along with timestamp, and actuator displacement. Sampling, recording, as well as remote control of Mach4 via RS232, was performed by an ESP32 on an Adafruit HUZZAH32 Feather board\footnote{\url{https://www.adafruit.com/product/3405}}.

Note that our tensile tester is able to move along three axes, which enables testing for shearing effects, however, this would require a modification for omni-directional force measurement. Although this would greatly complicate the procedure and evaluation, we see potential for future work, in order to simulate more generic actuation which may be closer to many real-life scenarios. Since this this work is focusing on variations of knit structures, we explicitly performed orthogonal actuation in our tests.

\begin{figure*}[!htb]
    \centering
    \textit{non-Lycra}\hspace{3in}\textit{Lycra}\vspace{5pt}\\
    \begin{subfigure}{.25\textwidth}
      \centering
      \includegraphics[width=1\textwidth]{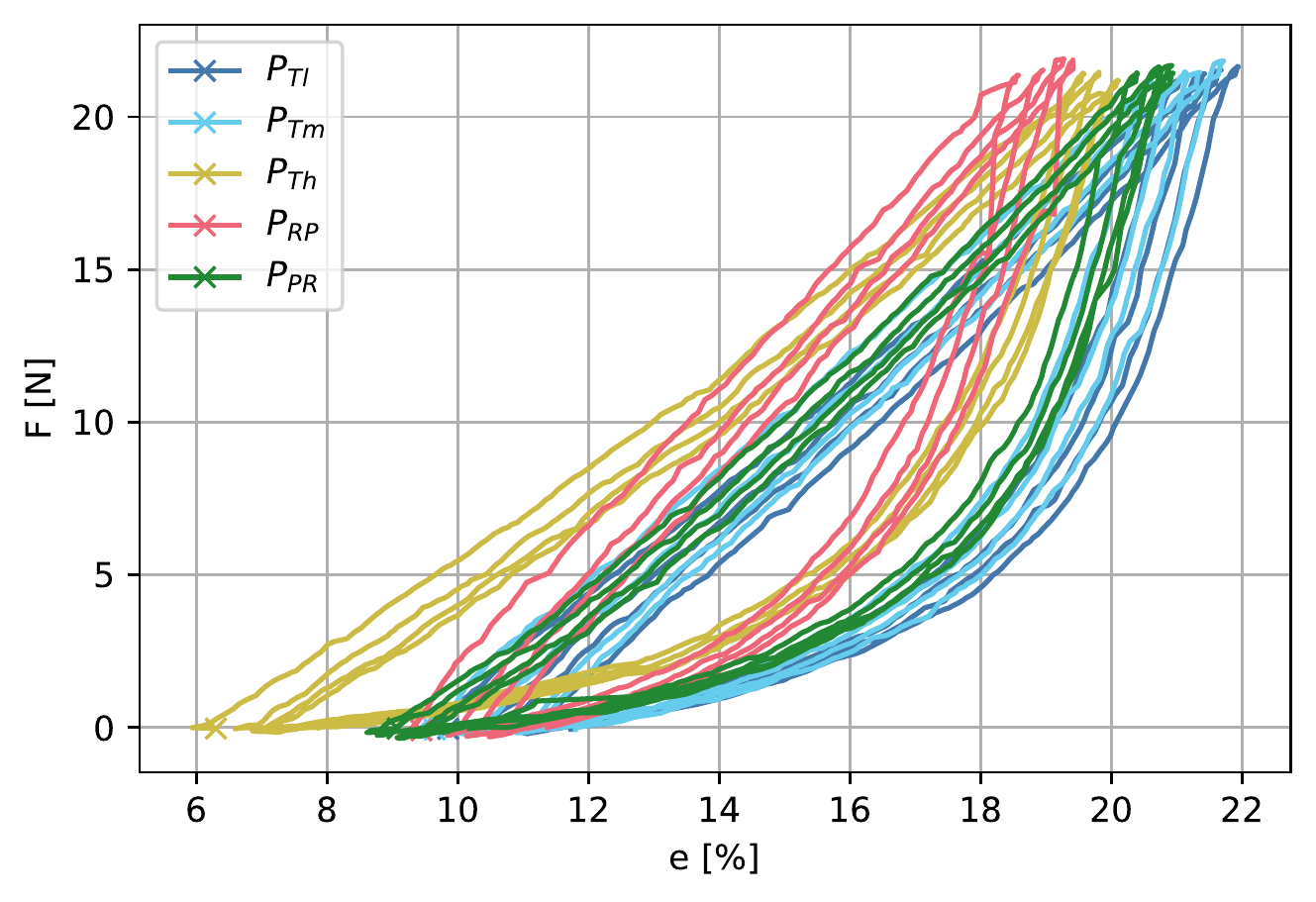}
      \caption{}
      \label{fig:characteristics:sub1}
    \end{subfigure}%
    \begin{subfigure}{.25\textwidth}
      \centering
      \includegraphics[width=1\textwidth]{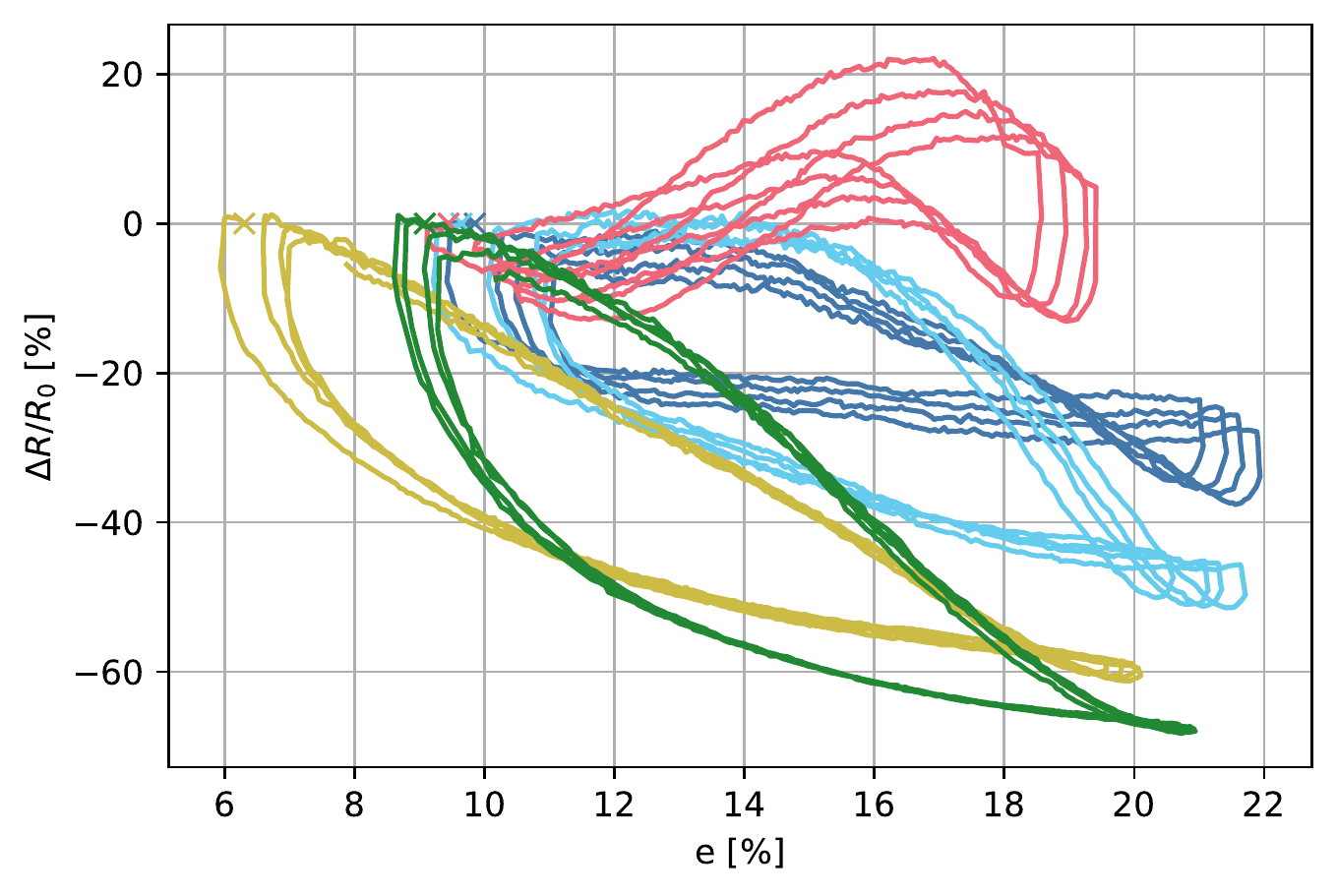}
      \caption{}
      \label{fig:characteristics:sub2}
    \end{subfigure}%
    \begin{subfigure}{.25\textwidth}
      \centering
      \includegraphics[width=1\textwidth]{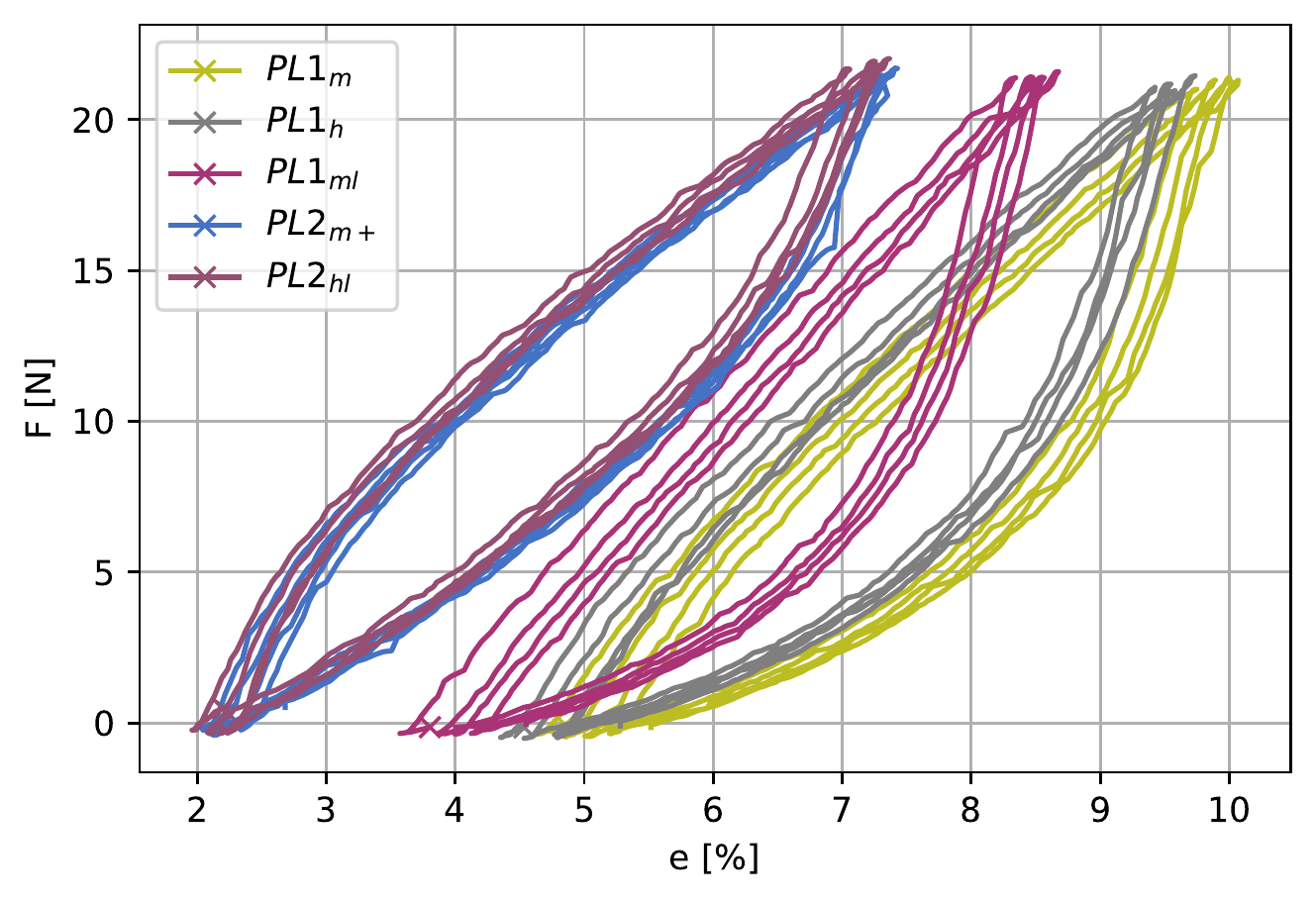}
      \caption{}
      \label{fig:characteristics:sub3}
    \end{subfigure}%
    \begin{subfigure}{.25\textwidth}
      \centering
      \includegraphics[width=1\textwidth]{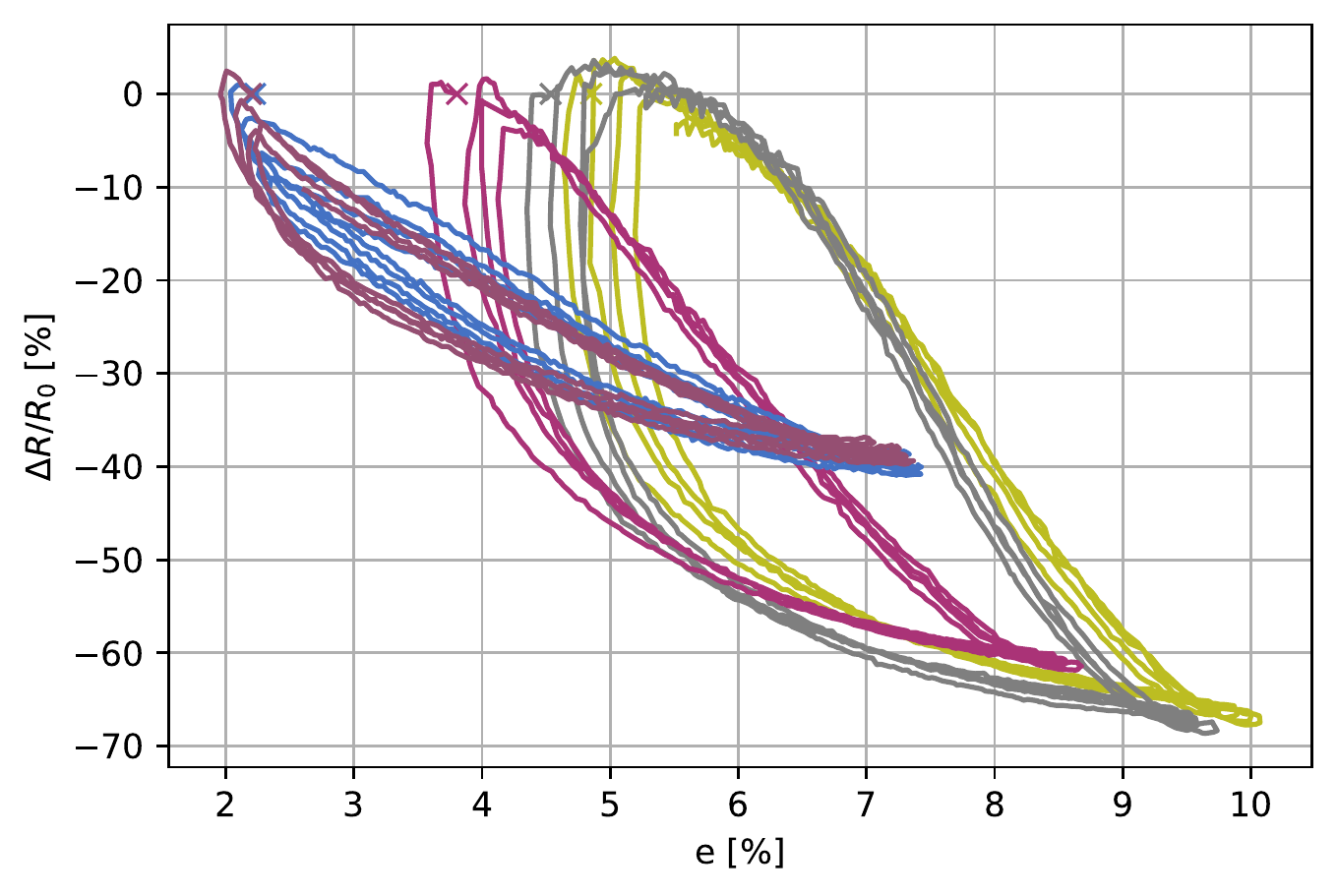}
      \caption{}
      \label{fig:characteristics:sub4}
    \end{subfigure}
    \begin{subfigure}{.5\textwidth}
      \centering
      \includegraphics[width=1\textwidth]{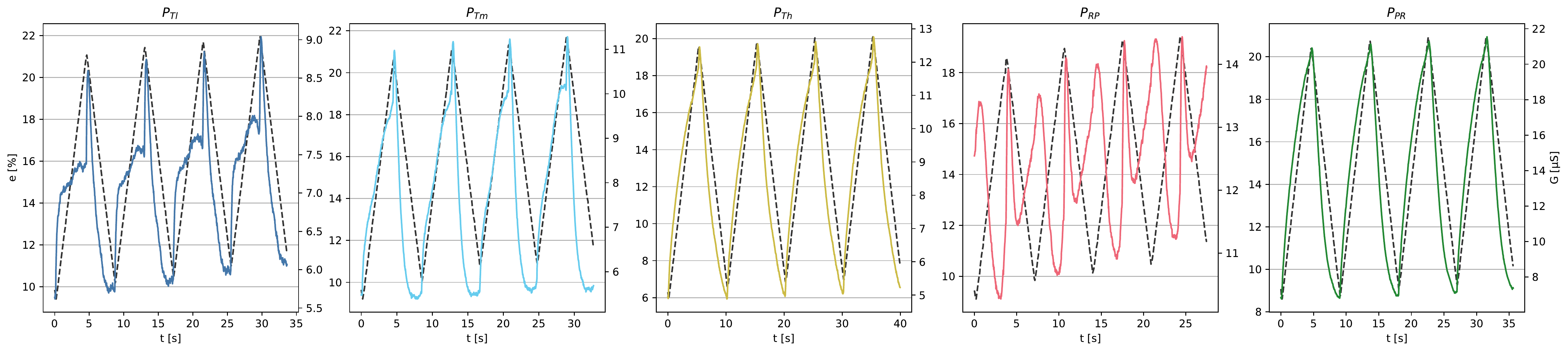}
      \caption{}
      \label{fig:characteristics:sub5}
    \end{subfigure}%
    \begin{subfigure}{.5\textwidth}
      \centering
      \includegraphics[width=1\textwidth]{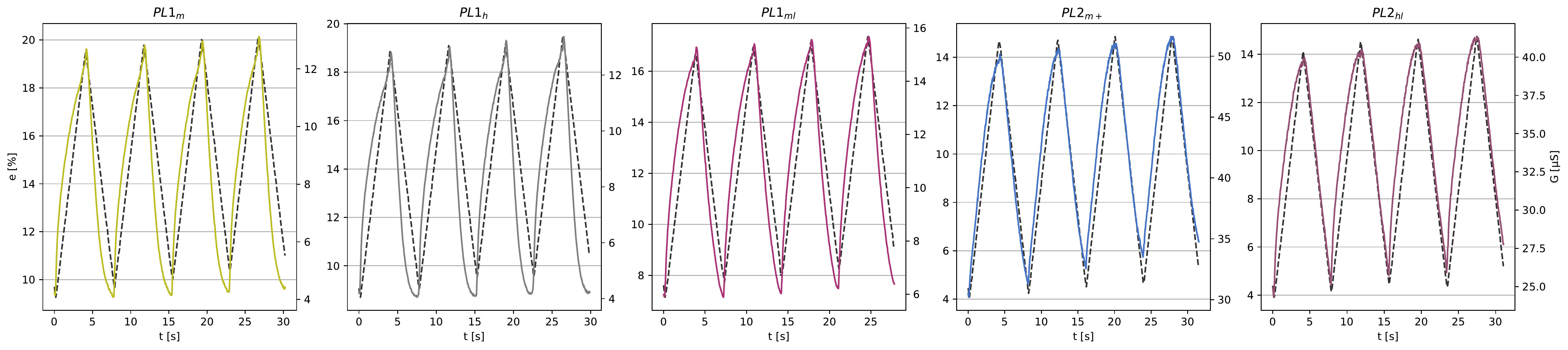}
      \caption{}
      \label{fig:characteristics:sub6}
    \end{subfigure}%
    
    \caption{Characteristics and timelines of non-Lycra (left) and Lycra (right) variants: Plots of correlation between strain $e$ and force $F$ (a,c), as well as strain and relative resistance change $\Delta R/R_0$ (b,d). 'x' marks initial values at beginning of recording. Timeline plots of all variations (e,f), overlaying strain $e$ (dashed, black) and sensor conductivity $G$, show respective conformity of our variations.}
    \label{fig:characteristics}
\end{figure*}

\subsection{Procedure}

All of our knitted sensor variations share a similar design (cf. Figure\,\ref{fig:sensor-tester} left), i.e., a 4\,cm $\times$ 4\,cm square field of resistive yarn, which is connected with conductive traces along the entire upper and lower courses. These connector traces were knit beyond the sensor area, leading to the edge of the textile sample, where we attached crocodile clips to connect our measurement electronics. We chose strong clips to avoid their slipping during the procedure and ensured adequate overlap with the conductive yarn. We refrained from testing different sensor dimensions, since we know from \cite{Ou2018} that sensor resistance is directly proportional to height and inversely proportional to width, with $R=\rho\frac{h}{w}$, where $\rho$ is a material-specific constant. We were able to verify this correlation in a preliminary evaluation (see supplementary material).

Each sensor variation was knitted three times. We performed an \textit{ex ante} evaluation to get an estimate regarding consistency and to identify outliers. We found good consistency overall and a very low number of outlier sensors (see supplement). We did however not perform an in-depth formal evaluation regarding consistency at this point.

For each sample, a single 5-cycle procedure was recorded. For the testing procedure, we marked the textiles at 5\,cm distance with the 4\,cm sensor areas centered, giving 5\,mm extra on each side for mounting. The samples were then punched through the mounting needles at the marks, so the tests would start from consistent initial lengths of 5\,cm. Tested patches were not previously ironed or otherwise chemically, mechanically, or thermally treated.

\subsubsection{Pulling with equal force}
To observe correlation of sensor reading and applied force, as well as sensor offset and general consistency, we performed a test procedure repetitively applying force along wale direction, and releasing again. We chose the force based on an informal initial test, where we estimated the upper working limit of most of our sensor variants with \mytilde 20\,N, and repeated for 5 cycles with a jog rate of 1.333\,mm/s. Note that, due to a communication lag in between the ESP32, Mach4, and the testing machine, we were slightly overshooting the target forces, however this does not undermine the general point of our results. Also note that, since different samples had different elastic behavior, this resulted in different strain ranges. Moreover, since we are not observing strain but force, we returned to F=0\,N after each cycle, which does not align with the initial actuator position of d=0\,mm, due to fabric extension from wear-out. As a result, this offset in strain could be considered a metric for the fabric's wear-out.

\subsubsection{Pulling with dwell}
In order to investigate drift and relaxation effects, we conducted a test similar to our initial one, however instead of switching actuation direction at 20\,N and 0\,N immediately, we dwelt for 5 seconds at each position. Note that due to ongoing fabric relaxation, the force was not constant at this point. We refrained from readjustment motions, since we judged this would introduce considerable jerkiness in the data and complicate analysis.

\subsubsection{Pulling with varying speed}
Since we noticed during \textit{ex-ante} experiments, that actuation speed can have a profound impact on the sensor reading -- most notably on the spikes in resistance after starting and stopping -- we repeated our initial tests (5 cycles at 20\,N, no dwell) with half and twice the baseline speed, hence, speeds were 0.667\,m/s, 1.333\,m/s, and 2.667\,m/s.

\subsubsection{Pulling with increasing force}
To find upper sensing range limit and to inspect consistency when pulled with different amplitudes, we varied pulling force. For this test, we started at an initial 5\,N and increased in steps of 5\,N up until 40\,N, with returning to 0\,N after each cycle. The test was again based on our initial one, i.e., we did not dwell before switching direction and moved with 1.333\,m/s.

\subsubsection{Long-term pull}
To observe long-term drift and relaxation effects, we performed a test pulling the samples to 20\,N and returning to 0\,N, dwelling for 15 minutes at each end.

\subsubsection{Course-directional pull}
In related work, knit strain sensors are frequently tested along a single direction \cite{Liang2019,Ayodele2021c,Bozali2021}. However, like most knitting structures such as Jersey and Rib patterns, a Twill is subject to anisotropic behavior in terms of physical properties, such as elasticity and recovery. We therefore investigated the behavior orthogonal to our primary testing direction as well, by mounting the sample rotated accordingly in our testing apparatus.

%% file: 04-results.tex
\begin{table*}
    \centering
    \caption{The majority of our results are gathered in this table. Note that we progressively excluded sensors that were performing badly from subsequent evaluation steps. Best values for non-Lycra and Lycra versions are put in bold.}
    \begin{tabular}{|r||c|c|c|c|c||c|c|c|c|c|}
        \hline
            & \multicolumn{5}{c||}{non-Lycra} & \multicolumn{5}{c|}{Lycra} \\
            & P\textsubscript{Tl} & P\textsubscript{Tm} & P\textsubscript{Th} & P\textsubscript{RP} & P\textsubscript{PR} & PL1\textsubscript{m} & PL1\textsubscript{h} & PL1\textsubscript{ml} & PL2\textsubscript{m+} & PL2\textsubscript{hl} \\
        \hline
        \hline
        $\Delta$d\textsubscript{0,5} {[\%]}    & 11.7  & 11.8  &  \textbf{7.9} & 11.4  & 10.2  & 11.0  & 10.6  & 9.1 & 5.4 & \textbf{5.2} \\
        $\Delta$d\textsubscript{1,5} {[\%]}     &  2.3  &  2.6  &  2.0 & 2.3 & \textbf{1.6} & 1.8 & 1.9 & 2.0 & \textbf{1.3} & \textbf{1.3} \\
        \hline
        $r^2$                             &  0.65 &  0.90 & \textbf{0.91} & -1.94 & 0.90 & 0.92 & 0.90 & 0.91 & 0.92 & \textbf{0.93} \\
        \hline
        h\textsubscript{R} {[\%]}               & 14.9  & 27.1  & \textbf{10.7}  &    - & 25.4  & 45.6 & 63.3 & 24.8 &  5.8  &  \textbf{4.1}  \\
        F\textsubscript{h} {[N]}                &  0.5  &  0.0  &  0.0  &    - &  0.0  &  0.0 &  0.0 &  0.0 &  9.2  & 11.1  \\
        \hline
        $\Delta$R\textsubscript{rel} {[\%]}     & 24.6 & 46.5 & 56.7 & - & \textbf{64.0} & 63.1 & \textbf{65.9} & 56.7 & 35.4 & 34.5 \\
        \hline
        offset {[\%]}                           & - & - & \textbf{-1.62} & - & -2.14 & -1.62 & - & - & -3.45 & \textbf{-3.24} \\
        relaxation {[\%]}                       & - & - &  7.32 & - &  \textbf{5.80} &    - &    - &    - &  2.59 &  \textbf{2.45} \\
        drift {[\%]}                            & - & - & \textbf{23.29} & - & 30.38 &    - &    - &    - &  8.51 &  \textbf{7.93} \\
        \hline
        T\textsubscript{r} [s]               & - & - & \textbf{22.9}  & - & 630.4 &    - &    - &    - &  \textbf{15.1}  &  \textbf{15.1} \\
        T\textsubscript{d} [s]               & - & - & 24.6  & - & \textbf{23.7} &    - &    - &    - & 10.7  & \textit{10.2} \\
        \hline
        jog x 0.5 $r^2$                   & - & - &  0.84 & - &  \textbf{0.94} &    - &    - &    - &  \textbf{0.92} &  0.20 \\
        jog x 2.0 $r^2$                   & - & - &  \textbf{0.90} & - &  0.87 &    - &    - &    - &  \textbf{0.94} &  0.84 \\
        \hline
        course-dir h\textsubscript{G} {[\%]}  & - & - &  \textbf{4.3}  & - &  8.7  &    - &    - &    - &     - &     - \\
        F {[N]}                                 & - & - &  6.9  & - &  5.5  &    - &    - &    - &     - &     - \\
        \hline
    \end{tabular}
    \label{tab:results}
\end{table*}

\section{Results and Discussion}
\label{sec:results}

In the following, we summarize our main findings. Note that to save space and reduce complexity, we narrow down our subset of evaluated patches, by progressively excluding poorly performing sensor variations.
Refer to Table \ref{tab:results}, which sums up the majority of our results.

\subsection{General performance}
\subsubsection{Non-Lycra variants}
Non-Lycra variants show almost linear relation between strain $e$ and applied force $F$ for the pulling segments (cf. Figure\,\ref{fig:characteristics:sub1}), however during release phase, we see considerable lag throughout all variations, which is due to poor elastic recoil. This effect inherently translates into hysteresis in sensor characteristics (cf. Figure\,\ref{fig:characteristics:sub2}), since releasing does not reflect in the knit mesh immediately and instead exhibits noticeable delay for recovery of the structure. We consider this an innate limitation of knits, however this can be counteracted to some degree, which was our main motivation to add Lycra into the supporting base knit, as outlined in Section \ref{subsec:variations}. 

We noticed that the resting state (i.e., at $F$=0\,N) of the sensor is highly different from the remaining iterations, hinting towards sensor offset. To quantify this wear-out effect, we calculated the relative extension $\Delta d_{0,5}$, i.e., the change in length in between \textit{before} 1\textsuperscript{st} and \textit{after} 5\textsuperscript{th} pulling iteration. We see that for the non-Lycra variants, P\textsubscript{Th} performs best, with 7.9\% extension. However, the first pulling iteration can be considered an outlier and may be irrelevant in many use case scenarios, e.g., when the fabric is draped and therefore permanently stretched. Therefore, we also report relative extension $\Delta d_{1,5}$, which excludes the first iteration by calculating relative length change in between \emph{after} 1\textsuperscript{st} and 5\textsuperscript{th} pulling iterations. There, the results are different, with P\textsubscript{PR} clearly outperforming P\textsubscript{Th}.

When comparing sensor response by relative change in resistance $\Delta R / R_0$ (cf. Figure\,\ref{fig:characteristics:sub2}), we see that for tubular structures (i.e., P\textsubscript{Tl}, P\textsubscript{Tm}, and P\textsubscript{Th}), tighter knit Resistat areas result in superior characteristics (i.e., P\textsubscript{Th}, with better correlation between strain and resistance change, less hysteresis, less noisy signal). Unexpectedly, there is considerable difference between the two connected variants P\textsubscript{RP} and P\textsubscript{PR}. While the patch with PES tucked to the Resistat (P$\rightarrow$R) clearly outperforms its tubular equivalent P\textsubscript{Tm}, tucking the other way around (P$\leftarrow$R) results in a defective sensor. We can eliminate the possibility of manufacturing flaws, since all of our three specimen of type P\textsubscript{RP} showed this erratic behavior. We exclude P\textsubscript{RP} from further evaluations, due to the bad performance.

The difference in sensor performance is also clearly visible on the timeline plots (cf. Figure\,\ref{fig:characteristics:sub5}), where conductivity $G$ of P\textsubscript{Th} and P\textsubscript{PR} goes well in line with strain $e$. Note that since we alternate between 0\,N and 20\,N, the values of $e$ drift slightly upwards due to wear-out effects. As mentioned above, we present strain-related data in the paper for sake of comparability, however we refer the interested reader to the supplement, which shows that the conductivity is well in line with the amplitude of force $F$ in most variations. We quantify conformance between the two trends of $F$ and $G$ using the Coefficient of Determination\footnote{We use the less-common lower case notation to mitigate confusion with electrical resistance $R$} $r^2=1-\textrm{SS}_{res}/\textrm{SS}_{tot}$, where $\textrm{SS}_{res}$ is the residual sum of squares and $\textrm{SS}_{tot}$ is the total sum of squares. Both data series are first normalized using the \texttt{preprocessing.StandardScaler} from Python package scikit-learn\footnote{\url{https://scikit-learn.org/stable/modules/generated/sklearn.preprocessing.StandardScaler.html}}, which transforms all values with $y = (x - \mu) / \sigma$. Results show that P\textsubscript{Tm}, P\textsubscript{Th}, and P\textsubscript{PR} perform best in that regard.

\subsubsection{Lycra variants}

From Figure\,\ref{fig:characteristics:sub3}, we see that elastic recoil was slightly improved by adding Lycra, however, the striking linearity we observed for the pulling-segments earlier seems to suffer from the boost in elasticity in general. For all of the Lycra variants PL*, conformity between force and conductivity is also slightly improved (cf. Figure\,\ref{fig:characteristics:sub6}, and $r^2$ values). Most notably the variants with 2$\times$Lycra (PL2*) show best linearity in e/R correlation with least hysteresis (cf. Figure \ref{fig:characteristics:sub4}). Drift also appears to be less severe for those variations, which also reflects on relative extension values $\Delta d_{0,5}$ and $\Delta d_{1,5}$ (cf. Table\,\ref{tab:results}). 

\subsection{Hysteresis \& Dynamic Range}

In order to objectively compare hysteresis, we separated data into pulling and releasing segments and fit exponential functions to the data sets using SciPy function \texttt{optimize.curve\string_fit}\footnote{\url{https://docs.scipy.org/doc/scipy/reference/generated/scipy.optimize.curve_fit.html}} (for further details refer to the supplementary material). We excluded the first pull/release cycle as an outlier for this curve fitting procedure and normalized $R$ values by scaling with $1/R_0$. We then searched for positions of maximum distance between pulling and releasing curves. Results are reported in Table \ref{tab:results}, with resistance hysteresis $h_R$, at respective locations $F_h$. We see that the variants with two Lycra threads PL2* outperform all others, including non-Lycra knits. Furthermore, their maximum hysteresis is found at $F$\mytilde 10\,N, unlike most patches, which show considerable differences at 0\,N, as results of strong settling effects.

We report dynamic range as relative difference in sensor resistance in between $F$=0\,N and $F$=20\,N, i.e., $\Delta R=\mathrm{abs}\left(R_0 - R_{20}\right) / R_0$. To find $R_0$ and $R_{20}$, we used the curves fit for finding hysteresis, and evaluated them at $F$=0\,N and $F$=20\,N: for non-Lycra versions, the connected sensor shows higher range than the tubular knits. The versions with two strands of Lycra PL2* show least range.

Even though the PL2* variants were inferior in terms of range, we decided to exclude PL1* sensors for further evaluation, since we judged low hysteresis and better consistency far more important than range. We furthermore tried to keep a balanced set of Lycra and non-Lycra patches with the two best-performing each.

\begin{figure*}[!h]
    \centering
    \includegraphics[width=1\textwidth]{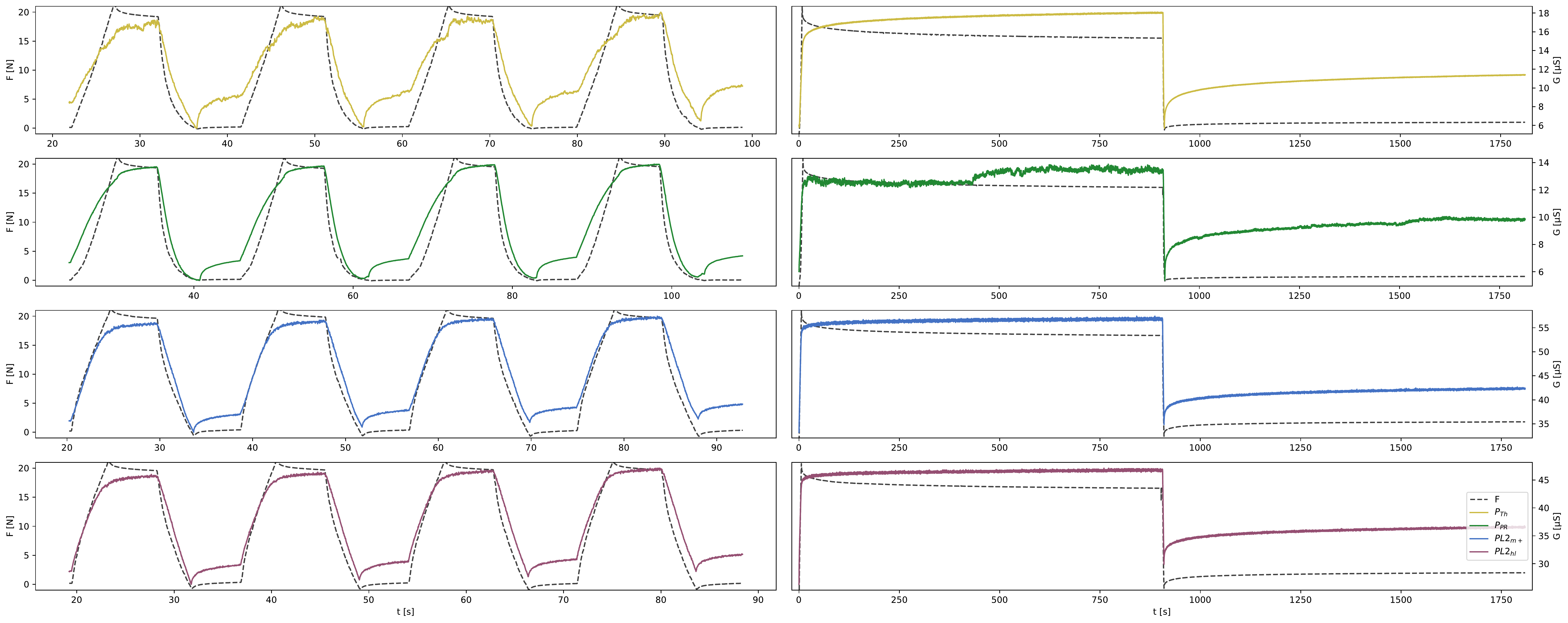}
    \caption{Results from short-term dwell with repetitive actuation (left) and long-term dwell with single actuation (right) show all sensors are subject to relaxation and sensor drift.}
    \label{fig:dwell-combined}
\end{figure*}

\begin{figure*}[!ht]
    \centering
    \begin{subfigure}{.5\textwidth}
        \centering
        \includegraphics[width=1\textwidth]{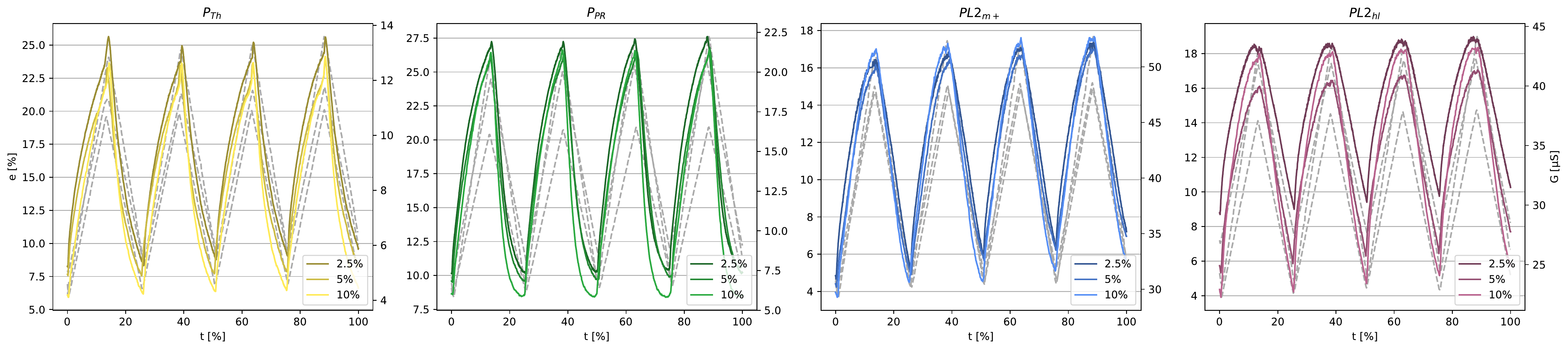}
        \caption{}
        \label{fig:jog-cascade:sub1}
    \end{subfigure}%
    \begin{subfigure}{.5\textwidth}
        \centering
        \includegraphics[width=\textwidth]{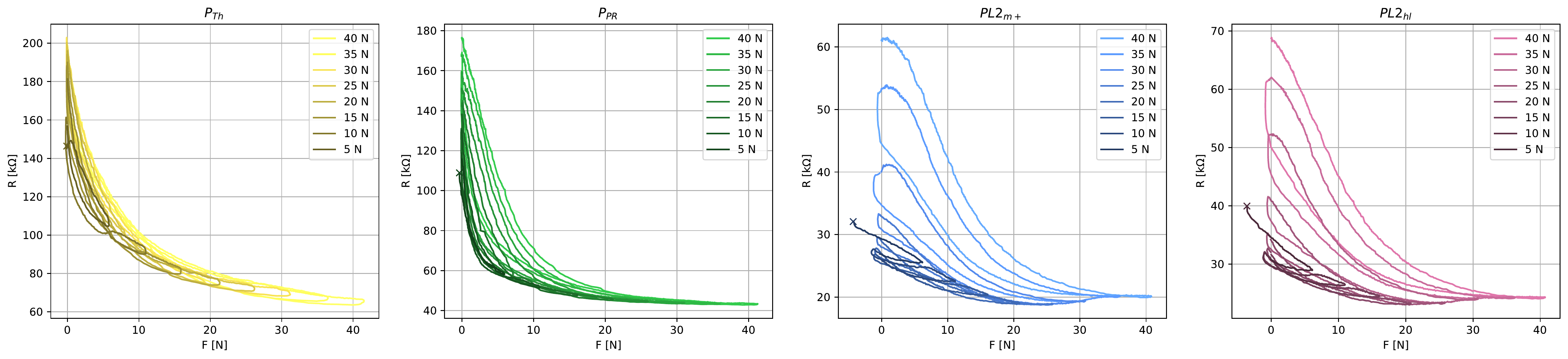}
        \caption{}
        \label{fig:jog-cascade:sub2}
    \end{subfigure}
    \caption{Results of our jog-rate test (a) showed that PL2\textsubscript{m+} is most immune against variations in actuation speed. Cascading tests (b) showed that characteristics lines of Lycra-variants follow largely different trends, depending on the force applied.}
    \label{fig:jog-cascade}
\end{figure*}

\subsection{Offset, Relaxation, \& Drift}
We use the terminology of \cite{Jansen2020}, with \textit{offset} being the change in resting-state resistance after each pulling iteration, \textit{relaxation} being the continuous change in resistance at constant strain, and \textit{drift} being the continuous change in resistance when returned to resting state. We calculate offset as relative change in resistance between \textit{before} and \textit{after} each cycle. To quantify relaxation and drift, we calculated resistance change relative to the initial value of the respective dwelling segment. For calculating all three metrics, we again excluded the first pull/release iteration for each sets, since sensors initiate form long-term settled states (cf. initial trends in Figures \ref{fig:characteristics:sub5} and \ref{fig:characteristics:sub6}); averaged values of the remaining 4 segments are presented in Table \ref{tab:results}.

Offset values show the remaining non-Lycra variants are superior over our Lycra versions. Furthermore, the tubular structure P\textsubscript{Th} outperforms the connected version P\textsubscript{PR}. However, tests with dwelling at 0\,N and 20\,N for 5 seconds each showed that Lycra variants are by far superior in terms of both relaxation and drift, as can be seen also in Figure \ref{fig:dwell-combined} (left). For Lycra variants, PL2\textsubscript{hl} is only slightly superior, while non-Lycra versions tie for both metrics: the tubular sensor shows less drift but worse relaxation behavior. 

Results from long-term test (cf. Figure \ref{fig:dwell-combined}, right) show similar results. It is clearly visible that P\textsubscript{PR} exhibits highest relative noise and P\textsubscript{Th} shows lowest. To compare settling of conductivity values, we calculated RSD over time windows of the past 10 seconds with $\textrm{RSD}_{10}(t) = \sigma_{10}(t) / \mu_{10}(t)$, where $\sigma_{10}$ and $\mu_{10}$ denote the SD and arithmetic mean of $G$ values in the period $[t-10\textrm{s}, t]$ respectively. In Table \ref{tab:results}, we specify periods for RSD\textsubscript{10} to \textit{permanently} drop below 1\%. We report one value for relaxation (i.e., when the sensor patch is pulled, T\textsubscript{r}) and one for drift (i.e., after sensor patch is released again, T\textsubscript{d}). Similar to our short-term dwelling tests, we observed that both Lycra variants are by far superior for both of relaxation and drift. Furthermore, the Lycra-versions' advantage is backed by long-term actuation tests, straining the sensors 2,200 times with e=20\% over a time-span of 5.6 hours (see supplement).

\subsection{Actuation speed}
To compare recordings with different lengths resulting from different jog-rates, we again discarded the first pull/release iterations as outliers, normalized along time-axes, and downsampled our data to equal sample count with \texttt{scipy.interpolate.interp1d}\footnote{\url{https://docs.scipy.org/doc/scipy/reference/generated/scipy.interpolate.interp1d.html}}. We again scaled our samples using \texttt{sklearn.preprocessing.StandardScaler} with $y = (x - \mu)/\sigma$, this time not individually for each recording, but using $\mu$ and $\sigma$ of our baseline set (1.333\,mm/s) for all three speeds, to preserve relative deviations. We then determined conformity of half and double speed from the baseline set by calculating $r^2$, which can be found in Table \ref{tab:results}. 

We see that for the non-Lycra versions, P\textsubscript{Th} deviates more for half speed, however less for double speed. For Lycra patches, we see clearly that PL2\textsubscript{m+} is superior to PL2\textsubscript{hl} (cf. Figure \ref{fig:jog-cascade:sub1}), which implies that tighter tuck tension (i.e., tighter connection between front and back faces) should be preferred.

\subsection{Increasing force}
Figure \ref{fig:jog-cascade:sub2} shows data collected from continuously increasing force for each iteration in steps of 5\,N. We see that both Lycra variants perform very poorly and follow a different trend for each repetition, returning to largely different sensor values when released. Note that this is most severe when we went beyond our previous upper testing limit of 20\,N. Non-Lycra patches show much more consistent trends, with P\textsubscript{Th} achieving the best results, even in highly-strained states. We see furthermore, that P\textsubscript{PR} seems to reach saturation more quickly.

\subsection{Anisotropy}

\begin{figure}
    \centering
    \begin{subfigure}{.5\columnwidth}
      \centering
      \includegraphics[width=\linewidth]{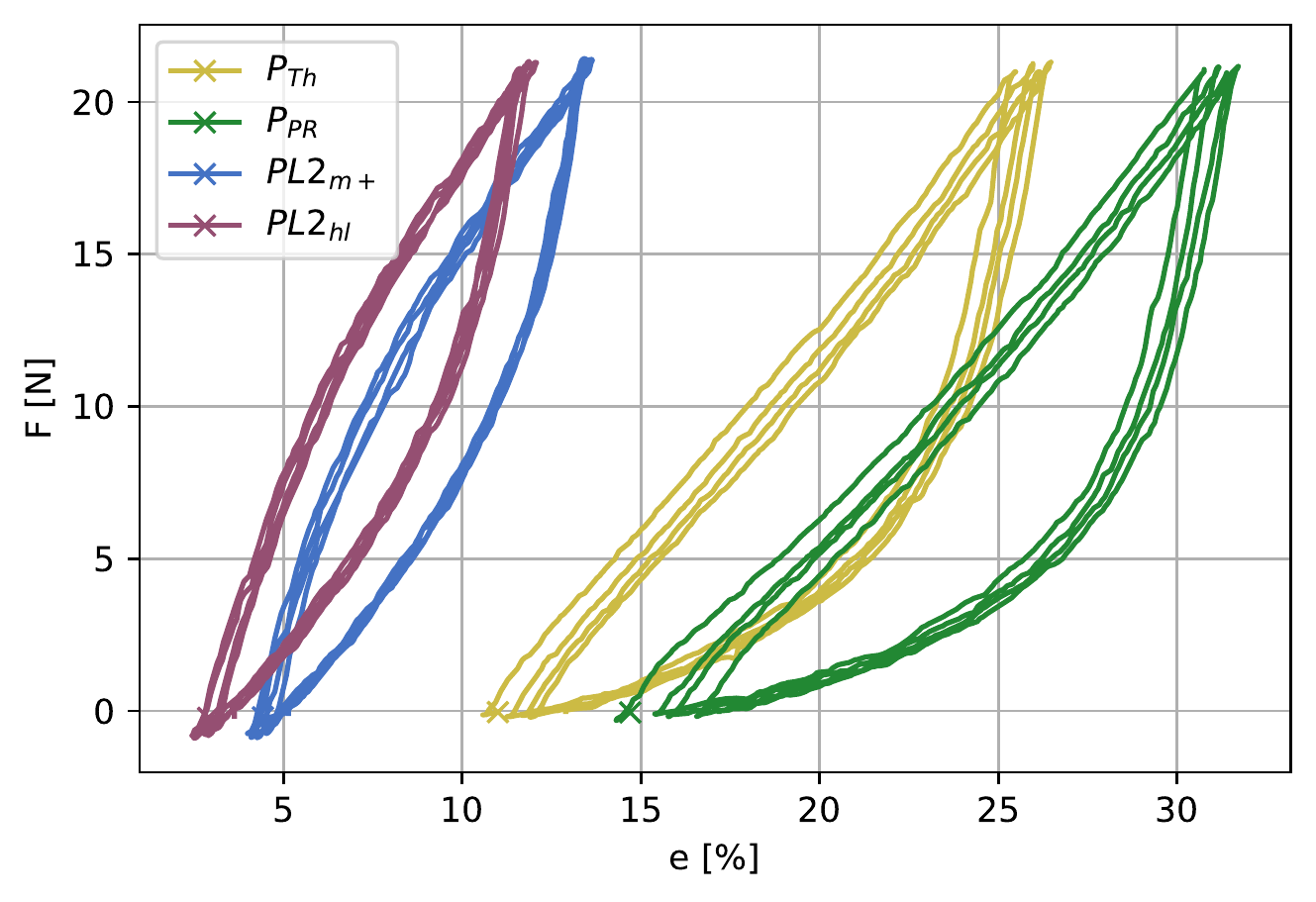}
      \label{fig:anisotropy:sub1}
    \end{subfigure}%
    \begin{subfigure}{.5\columnwidth}
      \centering
      \includegraphics[width=\linewidth]{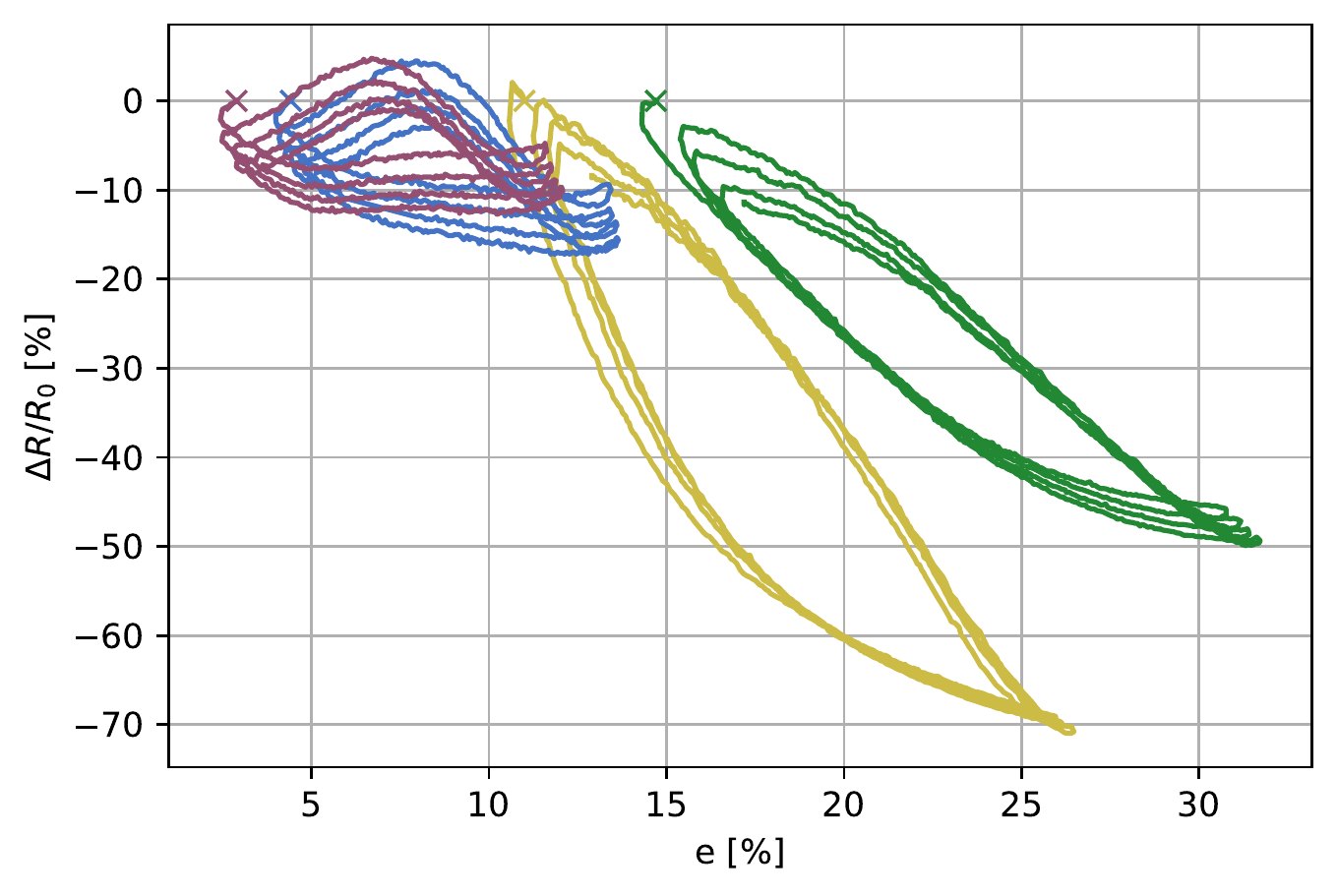}
      \label{fig:anisotropy:sub2}
    \end{subfigure}
    \begin{subfigure}{\columnwidth}
      \centering
      \includegraphics[width=\linewidth]{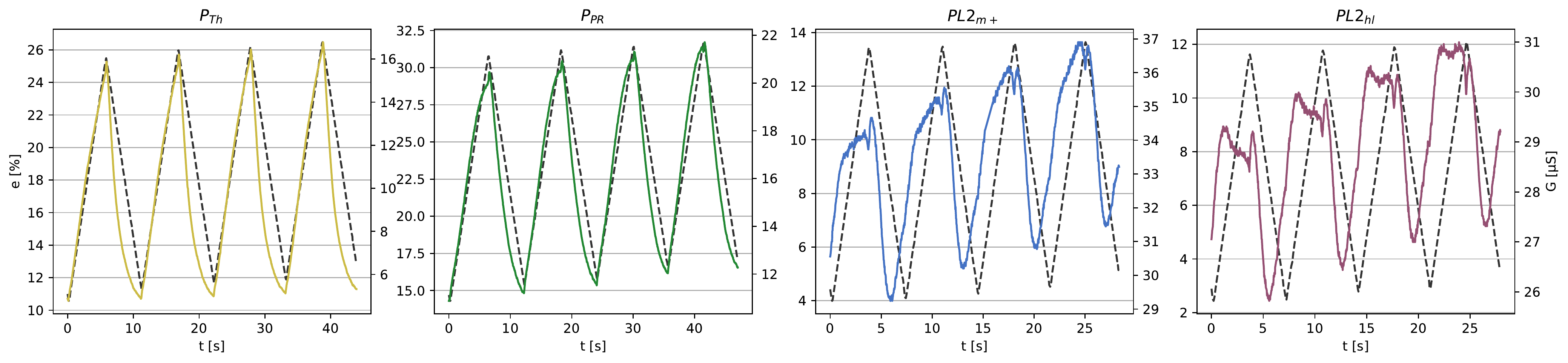}
      \label{fig:anisotropy:sub3}
    \end{subfigure}
    \caption{Course-directional strain behavior is most consistent for non-Lycra variants, while Lycra-versions act highly erratic.}
    \label{fig:anisotropy}
\end{figure}

Our main testing direction so far was along wales, where the knit provides good stability. Many use cases cause actuation along a single direction, however, given the anisotropic nature of most knitting structures, it is reasonable to also consider the orthogonal direction. Although for our Twill structure equal force causes higher strain in course-direction, first observations with non-Lycra patches did not hint towards considerable differences in terms of e/R correlation. Our tests however show that the Lycra-variants exhibit significant erratic behavior when actuated along courses. In particular, characteristics and timelines in Figure \ref{fig:anisotropy} show erratic sensor values for Lycra-variants, similar to P\textsubscript{RP}, which we discarded earlier. We therefore infer that patches with Lycra additives are only of limited value for use cases that involve omni-directional strain and/or shear. Hysteresis data for non-Lycra patches can be found in Table \ref{tab:results}. As earlier with P\textsubscript{RP}, we refrained from calculating hysteresis for Lycra-variants since we found curve fitting unreasonable due to their overall erratic behavior and high offsets in between pulling iterations; more details on curve fitting and hysteresis can be again found in the supplementary material.

\subsection{Discussion}

Summarizing, we found that when manufacturing our sensors as tubular knits, the resistive part should exhibit slightly higher tension than the substrate. Connecting front and back faces can yield better results in many instances, however an unexpected finding was that the tucking direction is crucial: if the resistive yarn is tucked to the substrate, the resulting sensor is defective. If however the tucks are performed the other way around, the result is among the best-performing sensors, however gives noisier values when compared to the tubular versions. Lycra-variants produced more consistent results overall, at least within our testing range of 0 to 20\,N. They subjectively showed less wear-out effects and better elastic recoil. In terms of quantitative measures, they are less prone to relaxation and drift, both in short- and in long-term. Knit using two threads of Lycra for the substrate showed best linearity, least hysteresis, and are least affected by variations in actuation speed; using higher tuck tension yielded slightly better results.

Still, we see that the best choice of materials depends largely on the specific use case at hand. We noticed that Lycra-enhanced variants perform well for wale-directional strain, however, the results were sobering when we varied actuation amplitudes, especially beyond our usual testing range with huge offsets as there was considerable offset and hysteresis visible in the plots. In contrast, the PES-only-patches show much less anisotropic behaviour and could be used even beyond 20\,N, in particular the tubular version P\textsubscript{Th}.

%% file: 05-conclusion.tex
\section{Conclusion, Limitations, \& Future Work}
\label{sec:conclusion}

In this paper, we presented three means of fabrication for implementing a resistive force sensor in flatbed weft knitting machines with a minimum of two beds. The chosen knitting pattern enables to knit the sensing part on one bed entirely, which allows for combinations with a supporting substrate. Our method therefore provides the possibility to hide away the functional part for aesthetic and/or protective purposes. Based on these methods, we presented and evaluated 10 variations, 5 of which used a PES-only substrate material. The remaining 5 combined PES with Lycra to improve their physical properties. 

We do acknowledge a few limitations of this study: first, we did not evaluate all possible combinations of nominal stitch numbers (cf. Table \ref{tab:knitVariations}), instead we chose settings driven by subjective measures of quality. We did this to keep the number of patches reasonably low. We did however experiment with other compositions and plies which were moderately successful, and only presented the most relevant ones in this paper.

Second, the stitch numbers for PES differ slightly for PL1* patches, which seems like they are not objectively comparable. We justify this with the change in haptic quality when new material of a different type is introduced into the composition, which requires stitch numbers to be adjusted accordingly, hence, we went for comparable haptic quality. 

Third, we did not test all of our sensors to their full saturation, i.e., did not cover the entire working range. We did this since in our use case, 20\,N was way beyond the expected upper limit, however we noticed especially during the tests with increasing actuation force, that some variations perform badly beyond this value. This connects to a limitation of our Lycra-patches: offsets and hysteresis depending on actuation amplitude pose a challenge in general, since there seems to be temporal data (i.e., the degree of "past actuation") required to infer correct force and/or strain at all times. We plan to investigate this aspect in future work using a specially trained Artificial Neuronal Network to act as a special filter. First steps into this direction already yielded promising results for compensating those temporal effects.

Finally, tests in harsh environmental conditions, such as in highly humid and extreme temperature scenarios were not performed at this point. Related work showed that conductive polymer composites can be affected in particular by high humidity \cite{Cochrane2007}, however, we expect that since this is a property of the material, the knitting structure does not have a profound influence in that regard. Since the sensing parts are entirely replaceable by similar products and our work was focusing on structural compositions and consistency benefits from adding Lycra material, we refrained from evaluating the specific materials that we used for our implementation, since we expect our results would reasonably translate to arbitrary resistive and conductive yarn.

%% file: main.bbl
\begin{thebibliography}{10}

\bibitem{Spencer2001}
D.~J. Spencer, {\em Knitting Technology: A Comprehensive Handbook and Practical
  Guide}.
\newblock Woodhead Publishing, 3~ed., 2001.

\bibitem{Luo2023}
Y.~Luo, X.~Chen, X.~Li, H.~Tian, S.~Li, L.~Wang, J.~He, Z.~Yang, and J.~Shao,
  ``Heterogeneous strain distribution based programmable gated microchannel for
  ultrasensitive and stable strain sensing,'' {\em Advanced Materials},
  vol.~35, no.~2, p.~2207141, 2023.

\bibitem{Li2022}
S.~Li, X.~Chen, X.~Li, H.~Tian, C.~Wang, B.~Nie, J.~He, and J.~Shao,
  ``Bioinspired robot skin with mechanically gated electron channels for
  sliding tactile perception,'' {\em Science Advances}, vol.~8, no.~48,
  p.~eade0720, 2022.

\bibitem{Chen2020}
X.~Chen, J.~Shao, H.~Tian, X.~Li, C.~Wang, Y.~Luo, and S.~Li, ``Scalable
  imprinting of flexible multiplexed sensor arrays with distributed
  piezoelectricity-enhanced micropillars for dynamic tactile sensing,'' {\em
  Advanced Materials Technologies}, vol.~5, no.~7, p.~2000046, 2020.

\bibitem{Shao2019}
J.~Shao, X.~Chen, X.~Li, H.~Tian, C.~Wang, and B.~Lu, ``Nanoimprint lithography
  for the manufacturing of flexible electronics,'' {\em Science China
  Technological Sciences}, vol.~62, no.~2, pp.~175--198, 2019.

\bibitem{Atalay2015}
O.~Atalay, W.~R. Kennon, and E.~Demirok, ``Weft-knitted strain sensor for
  monitoring respiratory rate and its electro-mechanical modeling,'' {\em IEEE
  Sensors Journal}, vol.~15, no.~1, pp.~110--122, 2015.

\bibitem{Ayodele2021b}
E.~Ayodele, T.~Bao, S.~A.~R. Zaidi, A.~M.~A. Hayajneh, J.~Scott, Z.-Q. Zhang,
  and D.~McLernon, ``Grasp classification with weft knit data glove using a
  convolutional neural network,'' {\em IEEE Sensors Journal}, vol.~21, no.~9,
  pp.~10824--10833, 2021.

\bibitem{Li2021}
Y.~Li, X.~Miao, J.~Y. Chen, G.~Jiang, and Q.~Liu, ``Sensing performance of
  knitted strain sensor on two-dimensional and three-dimensional surfaces,''
  {\em Materials \& Design}, vol.~197, p.~109273, 2021.

\bibitem{Akerfeldt2015}
M.~Åkerfeldt, A.~Lund, and P.~Walkenström, ``Textile sensing glove with
  piezoelectric pvdf fibers and printed electrodes of pedot:pss,'' {\em Textile
  Research Journal}, vol.~85, no.~17, pp.~1789--1799, 2015.

\bibitem{Han2019}
X.~Han, X.~Miao, X.~Chen, G.~Jiang, and L.~Niu, ``Research on finger movement
  sensing performance of conductive gloves,'' {\em Journal of Engineered Fibers
  and Fabrics}, vol.~14, p.~1558925019887622, 2019.

\bibitem{Lee2021}
S.~Lee, Y.~Choi, M.~Sung, J.~Bae, and Y.~Choi, ``A knitted sensing glove for
  human hand postures pattern recognition,'' {\em Sensors}, vol.~21, no.~4,
  2021.

\bibitem{Grassi2017}
A.~Grassi, F.~Cecchi, M.~Maselli, M.~Röling, C.~Laschi, and M.~Cianchetti,
  ``Warp-knitted textile as a strain sensor: Characterization procedure and
  application in a comfortable wearable goniometer,'' {\em IEEE Sensors
  Journal}, vol.~17, no.~18, pp.~5927--5936, 2017.

\bibitem{Liang2019}
A.~Liang, R.~Stewart, and N.~Bryan-Kinns, ``Analysis of sensitivity, linearity,
  hysteresis, responsiveness, and fatigue of textile knit stretch sensors.,''
  {\em Sensors (Basel)}, 8 2019.

\bibitem{Lin2015}
X.~Lin and B.-C. Seet, ``A linear wide-range textile pressure sensor integrally
  embedded in regular fabric,'' {\em IEEE Sensors Journal}, vol.~15, no.~10,
  pp.~5384--5385, 2015.

\bibitem{Tohidi2018}
S.~Dinparast~Tohidi, A.~Zille, A.~P. Catarino, and A.~M. Rocha, ``Effects of
  base fabric parameters on the electro-mechanical behavior of piezoresistive
  knitted sensors,'' {\em IEEE Sensors Journal}, vol.~18, no.~11,
  pp.~4529--4535, 2018.

\bibitem{Wicaksono2022}
I.~Wicaksono, P.~G. Hwang, S.~Droubi, F.~X. Wu, A.~N. Serio, W.~Yan, and J.~A.
  Paradiso, ``3dknits: Three-dimensional digital knitting of intelligent
  textile sensor for activity recognition and biomechanical monitoring,'' in
  {\em 2022 44th Annual International Conference of the IEEE Engineering in
  Medicine \& Biology Society (EMBC)}, pp.~2403--2409, 2022.

\bibitem{Zhang2006}
H.~Zhang, X.~Tao, T.~Yu, and S.~Wang, ``Conductive knitted fabric as
  large-strain gauge under high temperature,'' {\em Sensors and Actuators A:
  Physical}, vol.~126, no.~1, pp.~129--140, 2006.

\bibitem{Bozali2021}
B.~Bozali, J.~J.~F. van Dam, L.~Plaude, and K.~M.~B. Jansen, ``Development of
  hysteresis-free and linear knitted strain sensors for smart textile
  applications,'' in {\em 2021 IEEE Sensors}, pp.~1--4, 2021.

\bibitem{Yang2009}
K.~Yang, G.-l. Song, L.~Zhang, and L.-w. Li, ``Modelling the electrical
  property of 1×1 rib knitted fabrics made from conductive yarns,'' in {\em
  2009 Second International Conference on Information and Computing Science},
  vol.~4, pp.~382--385, 2009.

\bibitem{Ehrmann2014}
A.~Ehrmann, F.~Heimlich, A.~Brücken, M.~Weber, and R.~Haug, ``Suitability of
  knitted fabrics as elongation sensors subject to structure, stitch dimension
  and elongation direction,'' {\em Textile Research Journal}, vol.~84, no.~18,
  pp.~2006--2012, 2014.

\bibitem{Chen2019}
X.~Chen, B.~Li, Y.~Qiao, and Z.~Lu, ``reparing polypyrrole-coated stretchable
  textile via low-temperature interfacial polymerization for highly sensitive
  strain sensor.,'' {\em Micromachines (Basel)}, 11 2019.

\bibitem{Ayodele2021c}
E.~Ayodele, S.~A.~R. Zaidi, J.~Scott, Z.~Zhang, M.~Hafeez, and D.~McLernon,
  ``The effect of miss and tuck stitches on a weft knit strain sensor,'' {\em
  Sensors}, vol.~21, no.~2, 2021.

\bibitem{Liang2022}
X.~Liang, H.~Cong, Z.~Dong, and G.~Jiang, ``Size prediction and electrical
  performance of knitted strain sensors,'' {\em Polymers}, vol.~14, no.~12,
  2022.

\bibitem{Jang2021}
S.~Jang, J.~Y. Choi, E.~S. Yoo, D.~Y. Lim, J.~Y. Lee, J.~K. Kim, and C.~Pang,
  ``Printable wet-resistive textile strain sensors using bead-blended composite
  ink for robustly integrative wearable electronics,'' {\em Composites Part B:
  Engineering}, vol.~210, p.~108674, 2021.

\bibitem{Teyeme2020}
Y.~Teyeme, B.~Malengier, T.~Tesfaye, and L.~Van~Langenhove, ``A fabric-based
  textile stretch sensor for optimized measurement of strain in clothing,''
  {\em Sensors}, vol.~20, no.~24, 2020.

\bibitem{DeRossi2002}
D.~De~Rossi, F.~Carpi, F.~Lorussi, A.~Mazzoldi, E.~Scilingo, and A.~Tognetti,
  ``Electroactive fabrics for distributed, conformable and interactive
  systems,'' in {\em SENSORS, 2002 IEEE}, vol.~2, pp.~1608--1613 vol.2, 2002.

\bibitem{Semjonova2022}
G.~Semjonova, A.~Davidovica, N.~Kozlovskis, A.~Okss, and A.~Katashevs, ``Smart
  textile sock system for athletes' self-correction during functional tasks:
  Formative usability evaluation,'' {\em Sensors}, vol.~22, no.~13, 2022.

\bibitem{Holm1967}
{"Holm, Ragnar"}, {\em {"Electric Contacts: Theory and Application"}}.
\newblock "Heidelberg": "Springer Berlin", 4~ed., 1967.

\bibitem{Li2010}
L.~Li, W.~M. Au, K.~M. Wan, S.~H. Wan, W.~Y. Chung, and K.~S. Wong, ``A
  resistive network model for conductive knitting stitches,'' {\em Textile
  Research Journal}, vol.~80, no.~10, pp.~935--947, 2010.

\bibitem{Liu2016}
S.~Liu, C.~Yang, Y.~Zhao, X.~ming Tao, J.~Tong, and L.~Li, ``The impact of
  float stitches on the resistance of conductive knitted structures,'' {\em
  Textile Research Journal}, vol.~86, no.~14, pp.~1455--1473, 2016.

\bibitem{Baribina2018}
N.~Baribina, A.~Oks, I.~Baltina, and P.~Eizentals, ``Comparative analysis of
  knitted pressure sensors,'' 05 2018.

\bibitem{Aigner2022}
R.~Aigner, M.~A. Haberfellner, and M.~Haller, ``{SpaceR}: Knitting ready-made,
  tactile, and highly responsive spacer-fabric force sensors for continuous
  input,'' in {\em Proceedings of the 35th Annual ACM Symposium on User
  Interface Software and Technology}, UIST '22, (New York, NY, USA),
  Association for Computing Machinery, 2022.

\bibitem{Pointner2022}
A.~Pointner, T.~Preindl, S.~Mlakar, R.~Aigner, M.~A. Haberfellner, and
  M.~Haller, ``Knitted force sensors,'' in {\em Adjunct Proceedings of the 35th
  Annual ACM Symposium on User Interface Software and Technology}, UIST '22
  Adjunct, (New York, NY, USA), Association for Computing Machinery, 2022.

\bibitem{Isaia2020}
C.~Isaia, S.~A. McMaster, and D.~McNally, ``Study of performance of knitted
  conductive sleeves as wearable textile strain sensors for joint motion
  tracking,'' in {\em 2020 42nd Annual International Conference of the IEEE
  Engineering in Medicine \& Biology Society (EMBC)}, pp.~4555--4558, 2020.

\bibitem{Ou2018}
J.~Ou, D.~Oran, D.~D. Haddad, J.~Paradiso, and H.~Ishii, ``Sensorknit:
  Architecting textile sensors with machine knitting,'' {\em 3D Printing and
  Additive Manufacturing}, vol.~6, no.~1, pp.~1--11, 2019.

\bibitem{Jansen2020}
K.~M. Jansen, ``Performance evaluation of knitted and stitched textile strain
  sensors,'' {\em Sensors}, vol.~20, no.~24, 2020.

\bibitem{Cochrane2007}
C.~Cochrane, V.~Koncar, M.~Lewandowski, and C.~Dufour, ``Design and development
  of a flexible strain sensor for textile structures based on a conductive
  polymer composite,'' {\em Sensors}, vol.~7, no.~4, pp.~473--492, 2007.

\end{thebibliography}


\begin{thebibliography}{1}

\bibitem{Aigner2023}
R.~Aigner and F.~Hepper, ``An evaluation of multi-component weft-knitted twill
  structures for sensing tensile force,'' 2023.

\bibitem{Ou2018}
J.~Ou, D.~Oran, D.~D. Haddad, J.~Paradiso, and H.~Ishii, ``Sensorknit:
  Architecting textile sensors with machine knitting,'' {\em 3D Printing and
  Additive Manufacturing}, vol.~6, no.~1, pp.~1--11, 2019.

\end{thebibliography}
